\documentclass[epj,final]{svjour}
\usepackage{graphicx}
\usepackage{dcolumn}
\usepackage{bm}

\newcommand{\To}{\longrightarrow}

\newcommand{\A}{\mathbf{A}}
\newcommand{\G}{\mathbf{G}}

\newcommand{\I}{\mathbf{I}}
\newcommand{\B}{\mathbf{B}}
\newcommand{\M}{\mathbf{M}}

\newcommand{\Y}{\mathbf{Y}}
\newcommand{\W}{\mathbf{W}}

\newcommand{\PPP}{\mathbf{P}}
\newcommand{\Q}{\mathbf{P}^{-1}}
\newcommand{\e}{\mathbf{e}}

\newcommand{\x}{\mathbf{x}}

\newcommand{\uu}{\mathbf{u}}
\newcommand{\vv}{\mathbf{v}}
\newcommand{\0}{\mathbf{0}}
\newcommand{\LL}{\lambda}
\newcommand{\lb}{\langle}
\newcommand{\rb}{\rangle}

\begin{document}

\title{Moment instabilities in multidimensional
systems with noise}

\author{Dennis M. Wilkinson\inst{1,2}}

\institute{HP Labs, 1501 Page Mill Rd, Palo Alto, CA 94304 USA
\email{dennisw@hpl.hp.com} \and Physics Department, Stanford
University, 382 Via Pueblo Mall, Stanford, CA 94305-4060 USA}

\date{\today}


\abstract{We present a systematic study of moment evolution in
multidimensional stochastic difference systems, focusing on
characterizing systems whose low-order moments diverge in the
neighborhood of a stable fixed point. We consider systems with a
simple, dominant eigenvalue and stationary, white noise. When the
noise is small, we obtain general expressions for the approximate
asymptotic distribution and moment Lyapunov exponents. In the case
of larger noise, the second moment is calculated using a different
approach, which gives an exact result for some types of noise.  We
analyze the dependence of the moments on the system's dimension,
relevant system properties, the form of the noise, and the
magnitude of the noise. We determine a critical value for noise
strength, as a function of the unperturbed system's convergence
rate, above which the second moment diverges and large
fluctuations are likely. Analytical results are validated by
numerical simulations. We show that our results cannot be extended
to the continuous time limit except in certain special cases.
\PACS{{02.50.Ey}{Stochastic processes} \and {02.50.Sk}{Multivariate analysis}
\and {05.45.Ca}{Noise}}
}


\maketitle


\section{Introduction} \label{s:introduction_moment_instab}

\subsection{Motivation and previous work} The stability of fixed
points in a multidimensional system is easily ascertained when the
system is perfectly deterministic by using linear stability
analysis\cite{Guckenheimer}. Many real-world systems, however, are
not perfectly deterministic because their interactions are subject
to noise \cite{Gardiner}. It is therefore of interest to consider
the effect of a multiplicative noise term on a linearized system:
\begin{equation}\label{intro1}
  \x^t = (\A + \B^t)\x^{t-1}.
\end{equation}

In this paper we analyze the effect of white, stationary mean 0
noise in discrete systems. This type of noise has no effect on a
system's stability in mean, because the expected value evolves
exactly as if the system were unperturbed (\S
{\ref{s:expectedvalue}). However, multiplicative noise processes
cause fluctuations which can be large even if the fixed point is
stable (figure \ref{f:fluct}), knocking the system out of the
linear regime and coupling it to nonlinearities. Even for exact
linear models, large fluctuations can cause long delays in
convergence. An example of fluctuations in such a system is shown
is figure \ref{f:fluct}, where the dotted line shows evolution of
the first component of $\x$ without noise, and the solid line
shows one instance of evolution with noise.
\begin{figure}[!h]
     \centering
     \includegraphics[width=3in]{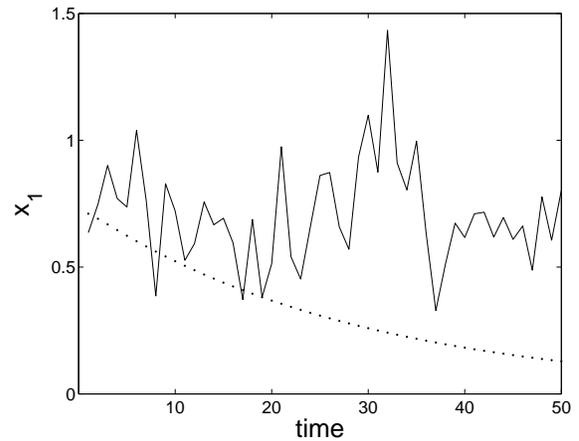}
     \caption{Example of fluctuations in a linear system}
     \label{f:fluct}
\end{figure}

Fluctuations in a stochastic system are studied by way of the
system's moments \cite{Gardiner}. The $p$th moment of a
multivariate system is simply the expected value of $|\x|^p$;
large moments, especially the low order moments such as the second
and third, indicate that a system attains large values with
non-negligible probability \cite{Rice}. For example, the system of
figure \ref{f:fluct} has a divergent moments for $p \geq 3$.

Multiplicative noise causes fluctuations because its effect is to
cause the moments of a system to diverge, even when the system
converges in mean \cite{Redner,LewontinCohen}. In particular,
divergent low-order moments in the neighborhood of a stable fixed
point are likely to cause the large fluctuations described above.
The evolution of the moments is thus an important consideration in
regards to fixed point stability in systems whose interactions are
subject to noise.

The asymptotic behavior of a random system and its moments is
characterized by the system's Lyapunov exponent and moment
Lyapunov exponents \cite{Arnold1986}. Calculation of Lyapunov
exponents for multivariate systems is very difficult in general,
even in simple cases \cite{Tsitsiklis}. Stability analysis and
calculation of Lyapunov exponents for discrete linear stochastic
systems and random matrix products has been a major area of
research in mathematics
\cite{BougerolLacroix,Stewartmatrixpowers,CohenNewman,Gurvits,Barabanov,Hinrichsenpencils,Hinrichsenpositive},
control theory \cite{KumarVaraiya,Xiao,Moustakidis,MiliVerriest},
physics \cite{CrisantiPaladin,VanKampen}, engineering mechanics
\cite{MaCaughey,Kozin,Haddad}, and biology
\cite{stabilityofecosystems,LewontinCohen}, among others. However,
the subject of most research has been stability in mean, not
stability of the moments. The traditional approach to determining
convergence in random systems is to use bounds (above mathematics
references; see also \cite{Khas1966,Mao}, for example, for
continuous systems).

There has been little previous work on calculating exact
expressions for moment evolution in multidimensional stochastic
systems. One exception is \cite{Arnoldperturbation} for continuous
2-dimensional systems, the results of which are discussed in
context in \S \ref{s:continuouslimit}.

\subsection{Problem statement and notation} \label{s:notation}
We are studying a system evolving according to the difference
equation
\begin{equation}\label{main1}
\x^t =  (\A + \B^t) \x^{t-1},
\end{equation}
or
\begin{equation}\label{main2}
  \x^t = [\prod_{\tau=1}^t(\A + \B^\tau)]\x^0.
\end{equation}
Here $\x$ is the system state, a vector of random variables and
$\B^t$ is a matrix of white noise processes with mean 0. (That is,
$\lb B^t_{ij} \rb = 0$ and $ \lb B^{t}_{ij}B^{t'}_{ij} \rb \sim
\delta_{tt'}$.) The initial state $\x^0$ of the system is assumed
to be fixed. The eigenvalues of the matrix $\A$ are $\lambda_i$;
the largest eigenvalue $\lambda_1$ or simply $\LL$ is
simple\footnote{Simple eigenvalues have algebraic multiplicity 1
and thus only one associated eigenvector.} and dominant, that is,
$\LL > \lambda_i$ for all $i \neq 1$.

The system size is $n$. We define the mean of the $\A_{ij}$ to be
$a$, and the variance to be $\sigma_A^2$. In the \textit{mean
value approximation}, $\A \approx a\G$ where $\G$ is the matrix
whose elements are all 1. We will be diagonalizing $\A$ into the
form $\mathbf{P}\mathbf{\Lambda} \mathbf{P}^{-1}$, where
$\mathbf{\Lambda}$ is the diagonal matrix of eigenvalues of $\A$,
and
\begin{eqnarray}\label{uv}
  \uu   & = & \mathbf{P}_{i1} = \mbox{normalized right eigenvector} \\
  \nonumber & & ~~~~~~~~~~~~~\mbox{corresponding to } \LL; \\
  \vv   & = & (\mathbf{P}_{1i}^{-1})^T = \mbox{non-normalized left eigenvector} \\
  \nonumber & & ~~~~~~~~~~~~~~~~~\mbox{corresponding to } \LL.
\end{eqnarray}
so that $|\uu| = 1$. Note that $\vv \cdot \uu = 1$ and so $|\vv|
\geq 1$.

A vector $\x^t$ converges in mean if $\lb x^t_i \rb$ converges for
all $i$ . We use any typical definition for convergence. The
system's fixed point is stable whenever the system converges in
mean, because the initial state is irrelevant to convergence
\cite{Strang}.

We define the \textit{pth moment} of the system to be $\lb
|\x^t|^p \rb$. Moment convergence can be elegantly expressed in
terms of \textit{moment Lyapunov exponents}, discussed in
\cite{Arnold1986} and defined as
\begin{equation}\label{momlyap}
  L_p = \lim_{t \rightarrow \infty} \frac{\log \lb |\x^t|^p \rb}{t}.
\end{equation}
The asymptotic behavior of the $p$th moment is then given by $\lb
|\x^t|^p \rb \sim e^{tL_p}$ and the $p$th moment converges if
\[
L_p \leq 0.
\]

Finally, in the case that all the elements of the noise matrix
have the same variance $b^2$, we define the \textit{critical
value} $b_c^2$ to be the level of noise above which the second
moment diverges.

\subsection{Overview of results}

The central results of this paper are the approximations and exact
expressions describing the evolution of moments of the system
(\ref{main1}), in particular the second moment. The small noise
case is treated first using a perturbation approach. This approach
allows us to calculate the system's approximate asymptotic
distribution and moment Lyapunov exponents. For larger noises, an
iteration technique is presented which gives both small and large
noise results for the second moment. For certain types of noise,
the iteration method allows us to calculate the second moment
Lyapunov exponent exactly in any system. These results appear to
be the first general analytic results for the Lyapunov exponents
of discrete multivariate systems.

The analysis of this paper is valid in discrete systems with a
simple, dominant eigenvalue. The eigenvalue requirement is
satisfied by all nonnegative systems (see appendix
\ref{s:primitive}) and many arbitrary systems. Nonnegative
\cite{BermanNeumann} and positive
\cite{FarinaRinaldi,positivesystems} discrete systems arise in
Markov models, and the fields of biology, population models,
economics (input-output models), finance, and cooperative problem
solving, among others. Applications to arbitrary systems are too
numerous to list.

Particular results of this paper are as follows. First, we show
that in the small noise regime, the problem of approximating the
asymptotic probability distribution of a multidimensional system
reduces to the scalar case, which is trivial (\S
\ref{s:scalarcase}, \S \ref{s:expansion}, \S
\ref{s:perturbationLyap}). We thus obtain the expression
\begin{equation}\label{result1}
  \lb |\x^t|^p \rb = |\lb \x^t \rb|^pe^{t\varepsilon^2\frac{p(p-1)}{2} +
O(\varepsilon^4) }
\end{equation}
where $\lb \x^t \rb$ is the expected (unperturbed) value of the
system at time $t$, and $\varepsilon$ is a small parameter which
depends on the noise and is calculated for various forms of noise
(equation (\ref{eps}) and table \ref{t:varepsilonnoises}). This
approximation is justified by simulation (figures
\ref{f:L},\ref{f:sizedependence},\ref{f:wsizedependence}) and its
accuracy is discussed briefly in \S \ref{s:accuracy}.

In the case of larger noise, the iteration approach of section \S
\ref{s:iteration} presents a methodology for calculating the
second moment Lyapunov exponent to any degree of accuracy in any
system, provided the noise elements have the same variance. The
exact value of the Lyapunov exponent is expressed as the largest
eigenvalue of a matrix and its accuracy is justified in the
simulation of figure \ref{f:crazyA}.

It is shown that the results of the iteration technique agree with
(\ref{result1}) for small noise, and with the $\LL \rightarrow 0$
limit for large noise. It is also shown that all results agree
with the trivial scalar case discussed in \S \ref{s:scalarcase}.

While the unperturbed value of the system depends only on the
initial state and the dominant eigenvalue in the asymptotic limit,
the moments depends on other properties of the system including
the system size and the form of the noise. It is shown that
\begin{itemize}
 \item the
effect of a given level of noise can be magnified, in some cases
greatly, if the dominant eigenvalue of the unperturbed system is
ill-conditioned (\S \ref{s:multivariateproperties}, \S
\ref{s:perturbationLyap});
  \item the destabilizing effect of the noise is
damped as the number of independent components of noise increases
(``destructive interference'' of independent noises) (\S
\ref{s:ndependence}, figure \ref{f:sizedependence});
  \item the destructive interference of independent noises  is maximized in the mean value limit (\S \ref{s:notation})
and is mitigated by any deviation from this limit (\S
\ref{s:ndependence}, figure \ref{f:wsizedependence}).
    \item large noise (\S \ref{s:crit_large_noise}), or small noise in systems with a very
    ill-conditioned dominant eigenvalue (\S \ref{s:furtheriter}, figure
    \ref{f:crazyA}), almost certainly destabilizes the
    system.
\end{itemize}

We also present a discussion of the critical value for the noise
variance above which the second moment diverges and fluctuations
become a major consideration. This discussion is largely
restricted to the case in which all the noise elements have the
same variance for simplicity. We obtain the following expression
for the critical value:
\begin{equation}\label{result_bcrit}
  b_c^2 = \frac{1}{n^k +
  \frac{f_vf_u\LL^2}{1-\LL^2}},
\end{equation}
where $f_u$ and $f_v$ are parameters related to $\A$ and the type
of noise considered which are very close to 1 in the large
majority of systems, and $k$ equals 1 if all the noise elements
are independent and 2 if they are all correlated. This expression
is shown to be accurate in both the small and large noise cases,
and is used to create a stability diagram for the system in
figures \ref{f:phaseplotb} and \ref{f:phaseplotq}.

The dependence of the critical value on system parameters is
discussed. We show that
\begin{itemize}
  \item for small noise, the critical value depends weakly on the
  system size and type of noise considered
(\S \ref{s:crit_n_dependence}, figure \ref{f:bcritndependence});
 \item for large noise, the critical value depends strongly on the
  system size and type of noise
(\S \ref{s:crit_n_dependence});
  \item the critical value provides a much more accurate indication of the level
  of noise below which the second moment converges than a simple bound on
  convergence (\S \ref{s:crit_compare_to_bounds}, appendix \ref{s:bounds});
  \item for most convergent systems
 subject to small noise, the low-order moments diverge only if the unperturbed system converges
 slowly (\S \ref{s:crit_small_noise}).
\end{itemize}
This last statement is especially true for positive systems
(figure \ref{f:vgap}); note that systems with slow convergence may
have other problems besides fluctuations due to noise, such as
large transient behavior\cite{Trefethen}.

Finally, we consider the continuous limit and show that our
results only extend to this limit in certain very special cases
(\S \ref{s:mvacontinuous}, \S \ref{s:arnold}).

\subsection{Paper organization}
The paper is organized as follows. In \S \ref{s:preliminaries} we
present simple preliminary results: asymptotic expressions for the
system's expected value, moments in the scalar ($n=1$) case, and
second moment in the case that $\A=0$. Section
\ref{s:multivariateproperties} discusses the important properties
of the multivariate system. Moment evolution is calculated for
multivariate systems in the small noise limit using a perturbation
approach in \S \ref{s:perturbation}, and the result is discussed.
Section \ref{s:iteration} uses an iteration approach to treat the
second moment's evolution in the case of larger noise. The
critical value of noise for second moment divergence is the
subject of \S \ref{s:bcrit}. The accuracy of the approximations is
justified in numerical simulations throughout the paper. Finally,
\S \ref{s:continuouslimit} presents a discussion of the continuous
time limit.

\section{Preliminaries} \label{s:preliminaries}
This section presents a calculation of the expected value of the
system, as well as calculations of two limiting cases: a scalar
stochastic system, and a multivariate system with no fixed part
(noise only).

A discussion of the expected value of the system and its
convergence properties is a necessary preliminary step to any
study of the moments. The computation is trivial, as we show,
because the noise is white with mean zero.

Calculation of the moments of a scalar system provides a framework
which we will apply in the small noise limit of the multivariate
case (\S \ref{s:perturbation}). The scalar system also provides a
demonstration of how multiplicative noise leads to a log-normal
distribution and moment divergence.

The noise only, $\A = \0$ multivariate system is a system in which
the moments can be found exactly, yielding the zeroth order term
for the large-noise limit (see \S \ref{s:large_noise}). The
calculations involved also provide a useful preview of those in \S
\ref{s:iteration}.

All of the calculations in this section are quite straightforward
and have very likely been presented, in whole or part, in some
previous work. However, we did not find a specific reference with
the exception of \cite{MiliVerriest} for a cursory treatment of
the scalar case.

\subsection{Expected value and unperturbed system}
\label{s:expectedvalue} The expected (average) state of the system
and the state of the unperturbed system are equivalent since the
noise is white with mean 0. White noise means that $\x^{t-1}$ and
$\B^t$ are independent, so that
\[
 \lb \x^t \rb = (\A + \lb \B^t \rb) \lb \x^{t-1} \rb = \A \lb \x^{t-1} \rb,
\]
since the mean is 0. Thus
\[
\lb \x^t \rb= \A^t \x^0 = \x^t_{\textnormal{unperturbed}}.
\]
In systems with a simple dominant eigenvalue $\LL$, the asymptotic
behavior of unperturbed system is completely determined by the
largest eigenvalue of $\A$\cite{Strang}. For large $t$,
\begin{equation}\label{xb0}
  \x_{\textnormal{unperturbed}}^t = \LL^t (\vv\cdot \x^0)
  \uu
\end{equation}
and the moment Lyapunov exponents are simply
\begin{equation}\label{L0}
  L_p^0 = p \log \LL.
\end{equation}
The system will converge to 0 for any initial conditions if $\LL <
1$, and it will diverge if $\lambda_1>1$. In the case of
stochastic matrices with $\LL = 1$, the above formula is accurate
because $\LL$ is simple.  We are not interested in the case in
which $\x^0$ is orthogonal to $\vv$.

\subsection{Scalar stochastic system}
\label{s:scalarcase} In the case $n=1$ it is not difficult to
determine the asymptotic distribution, as well as the exact
expressions for any moment. We go through a derivation here
because this analysis will apply to the small noise multivariate
case. The scalar system is
\[
x_t = x_0\prod_{\tau=1}^t(a + b_\tau),
\]
where $\lb b_\tau \rb = 0$ and $\lb b_\tau b_{\tau'} \rb =
b^2\delta_{\tau\tau'}$. Notice that we express time as a subscript
in this section, whereas in the multidimensional treatment time is
a superscript.

\subsubsection{Exact expressions}
We have
\begin{eqnarray} \label{exactscalarmoments}
\lb x^p_t \rb &=& x_0^p \lb [a + b_\tau]^p \rb ^t \nonumber \\
 &=& x_0^pa^{pt}\left( \sum_{k=1}^p {p \choose k} \lb b_\tau/a \rb^k
 \right)^t.
\end{eqnarray}
In particular, $\lb x_t \rb= a^t$ and $\lb x^2_t \rb = (a^2 +
b^2)^t$, so that
\begin{equation}\label{scalar_L2_exact}
  L_2 = \log(a^2 + b^2).
\end{equation}

\subsubsection{Approximate asymptotic distribution}
\label{s:scalarapprox} In this subsection we assume small noise,
that is, $|b_\tau/a| < 1$. This allows us to take logs and ensures
that the moments of $b_\tau/a$ are well behaved. We have
\[
\log \frac{x_t}{x_0} = t \log a + \sum_\tau s_\tau
\]
where
\[
s_\tau = \log(1 + b_\tau/a).
\]
The $s_\tau$ are i.i.d., so the sum is normal for large $t$ with
mean $t\mu_s$ and variance $t\sigma_s^2$, where $\mu_s$ and
$\sigma_s^2$ are the mean and variance of the $s_\tau$, by the
central limit theorem. The system is thus log-normally distributed
in the asymptotic limit and its moments are given by
\begin{equation}\label{lnmom}
  \lb x_t^p \rb = x_0^p a^{pt} e^{pt\mu_s + p^2t\sigma_s^2/2}.
\end{equation}
Since we know that the first moment $\lb x_t \rb = x_0a^t$ is
independent of the noise, we can conclude that $\mu_s =
-\sigma_s^2/2$ and we have
\begin{equation}\label{scalarmom}
\lb x^p_t \rb= \lb x_t \rb^p e^{-t\mu_s p(p-1)}
\end{equation}
in the large $t$ limit. Thus the Lyapunov exponents are given by
\[
  L_p = p \log a - \mu_sp(p-1),
\]
or
\begin{equation}
    L_p = L^0_p - \mu_sp(p-1)
\end{equation}
where $L^0_p$ is the Lyapunov exponent for the unperturbed system.
Notice that $\mu_s < 0$ because the log function weights the
negative values of $b_t/a$ more heavily than the positive ones.

Expanding the log in the expression for $\mu_s = \lb \log(1 +
b_\tau/a) \rb$ we find:
\[
\mu_s = -\sum_k \frac{\lb b_\tau/a \rb^k}{k}(-1)^k.
\]
The $\lb b_\tau/a \rb^k$ term in the expansion must be $O(b/a)^k$
or smaller since $b_\tau/a$ can never exceed 1. Thus
\begin{equation}\label{Lpscal}
L_p = L^0_p  + p(p-1)(b/a)^2/2 + O(b/a)^3.
\end{equation}
The error is $O(b/a)^4$ if the noise is symmetric. In particular,
for the second moment,
\begin{equation}\label{L2scal}
  L_2 \approx L^0_p + (b/a)^2 + O(b/a)^3,
\end{equation}
in agreement with the exact value to second order.

The approximation and exact results for a scalar system are
compared to simulation in figure \ref{f:scalarmoments} below. This
system  converges in mean but has diverging moments for $p \geq
3$. The parameters for the simulation are $a = 0.97$ and normal
noise with $b^2 = 0.05$. The solid lines show the average of the
moments $\lb x_t^p \rb $ for $p = 1, 2, 3$ and 4, over $10^6$
runs. The dashed lines are the exact prediction
(\ref{exactscalarmoments}) and are shown only for $p = 3$ and 4.
The crosses are the approximation (\ref{Lpscal}); the inaccuracy
for $p=4$ is due to the expansion of the log. The initial value
was $x_0=1$ and noises larger than $a$ were not allowed.
\begin{figure}[htb]
     \centering
     \includegraphics[width=3in]{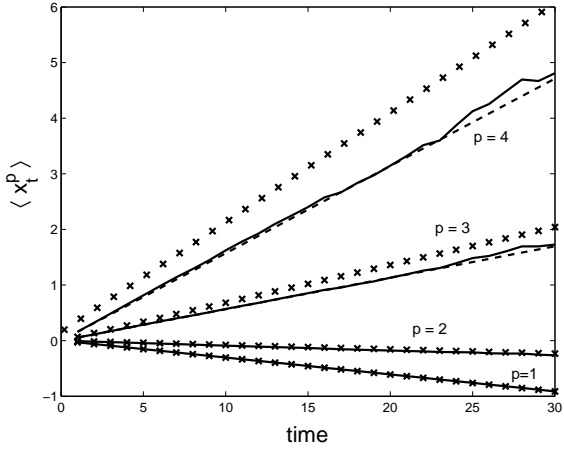}
     \caption{Moment evolution in scalar system}
    \label{f:scalarmoments}
\end{figure}

\subsection{$\A=\0$ limit of multivariate case} \label{s:Aequal0}
Returning to the multivariate system, we consider as a first
treatment the $\A = \0$ limit. The system's evolution in this
limit is given by
\begin{equation}\label{Aequal0main}
  \x^t =  \big[ \prod_{\tau=1}^{t} \B^{\tau} \big ] \x^0.
\end{equation}
The expected value of $\x$ is $\mathbf{0}$. The expression for
moments contains two or more occurrences of each $\B^{\tau}$; the
difficulty in its evaluation, and in general the difficulty of any
multivariate system, is that the noise matrices do not commute.
However, when $\A = \mathbf{0}$ and the noise is white, the sum
may be evaluated explicitly. Its value depends on the type of
noise considered. In this section we show the details of how to
evaluate such a sum; in later sections such steps will be skipped.

In principle any moment of $\x$ could be calculated exactly, given
a particular distribution for the noise elements. Here we restrict
our calculation to the second moment for clarity and simplicity.

\subsubsection{Independent noises} We first consider the case
where all the noise elements vary independently.  The matrices
$\B^\tau$ do not commute so we must consider the full term by term
expansion to evaluate the second moment:
\begin{eqnarray}
\lb |\x^t|^2 \rb & = &
 \sum_i  \sum_{j_{1}j_{2}\ldots
j_{t}}\sum_{k_{1}k_{2}\ldots k_{t}} \left\lb
B^{1}_{j_{1}j_{2}}B^{2}_{j_{2}j_{3}}\cdot \ldots \right. \\
 \nonumber & & ~~~~
\left.  \cdot B^{t}_{j_{t}i} x^{0}_i
B^{1}_{k_{1}k_{2}}B^{2}_{k_{2}k_{3}}\cdot\ldots\cdot
B^{t}_{k_{t}i} x^{0}_i \right\rb
\end{eqnarray}
where the expected value goes inside the sum because it is a
linear function. All the elements of every $\B$ are independent,
and we get a $\delta_{j_{\tau}k_{\tau}}$ for every $\tau = 1\ldots
t$ when we sum on the $k$'s. This gives
\begin{equation}\label{large_noise_general}
\lb |\x^{t}|^{2} \rb = \sum_{j_{1}j_{2}\ldots j_{t} i} \lb
(B^{1}_{j_{1}j_{2}})^{2}\rb \lb (B^{2}_{j_{2}j_{3}})^{2}\rb
\cdot\ldots\cdot \lb (B^{t}_{j_{t}i})^{2} (x^{0}_i)^{2}\rb
\end{equation}
When all the noise elements have the same variance $b^2$, each of
the $t$ sums on $j_{2},\ldots,j_{t}$ and $i$ simply gives a factor
of $nb^2$. The remaining sum is just the norm squared of $x^0$,
and we obtain
\begin{equation}\label{large_noise_UH}
\lb |\x^{t}|^2 \rb = (nb^2)^t |\x^0|^2.
\end{equation}

If the noises did not all have the same variance, the result would
be identical with $b^2$ replaced the average variance
\[
\bar{b}^2 = \frac{\sum_{ij} \lb (B_{ij})^2 \rb}{n^2}.
\]

\subsubsection{Correlated noises} When all the noises are
correlated with the same variance (T noise) the calculation is
similar except that both the sum on the $\{j_{\tau}\}$ and the sum
on the $\{k_{\tau}\}$ in equation (\ref{large_noise_general}) give
a factor of $n$. We thus obtain
\begin{equation}\label{large_noise_T}
\lb |\x^{t}|^2 \rb = (n^2 b^2)^t |\x^0|^2.
\end{equation}
We note that this result can be also found immediately by
transforming to the eigenspace, since in this case $\B$ is
proportional to the matrix $\G$ of all ones which has $\LL = n$
and all other eigenvalues equal to 0.

The case of noises where only certain elements are correlated
provides an intermediate case between independent and correlated
noises. The expressions are complex and are left for future work.

\bigskip

\begin{figure}[htb]
     \centering
     \includegraphics[width=3in]{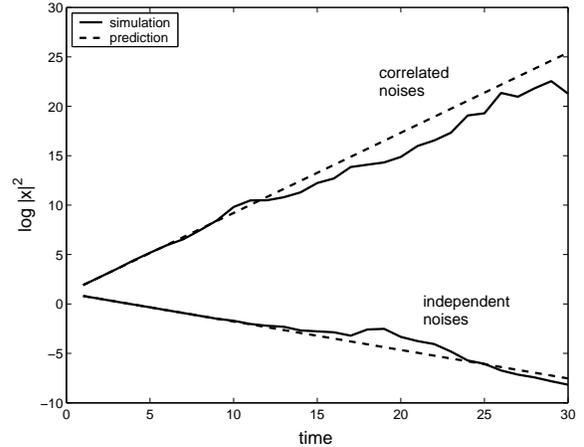}
     \caption{Second moment evolution for $\A=\0$}
    \label{f:Aequal0}
\end{figure}
Second moment evolution in the noise-only case is shown in figure
\ref{f:Aequal0}. The simulations show the value of $|\x|^2$
averaged over 1,000 runs, in the independent noise case, and
100,000 runs for correlated noise\footnote{A large number of runs
is necessary in the correlated noise case because of the divergent
moments \cite{Redner}}. Here $n=3$ and the elements of $\B$ were
chosen from a normal distribution with variance 0.25. The
simulations are compared to the predictions of
(\ref{large_noise_UH}) and (\ref{large_noise_T}).

\section{Properties of multivariate stochastic systems}
\label{s:multivariateproperties}
\subsection{Log-normal character of distribution}
\label{s:lognormal} As we saw in \S  \ref{s:scalarcase}, scalar
stochastic systems with stationary multiplicative noise are
log-normally distributed with parameters proportional to time, so
the system moments evolve as $\exp[tp(\mu + p\sigma^2/2)]$. While
$\mu$ is typically negative, for large $p$ the positive
$p\sigma^2/2$ term dominates and causes divergence. The effect of
the multiplicative noise is thus to cause the system's $p$th
moments to diverge for all $p$ greater than some $p_0$.

While the components of multidimensional stochastic systems with
multiplicative noise do not have an exact log-normal distribution,
they retain the general log-normal character including the heavy
tail and divergent moments. To be exact, any element of a product
of $t$ stationary random matrices is asymptotically log-normally
distributed with parameters proportional to
$t${\cite{Bellman,FurstenbergKesten}. Components of a multivariate
stochastic difference system are thus linear combinations of
log-normal variables with parameters proportional to $t$. Just as
in the scalar case, therefore, multiplicative noise in
multivariate systems causes the system's moments to diverge.

\begin{figure}[h!]
     \centering
     \includegraphics[width=3in]{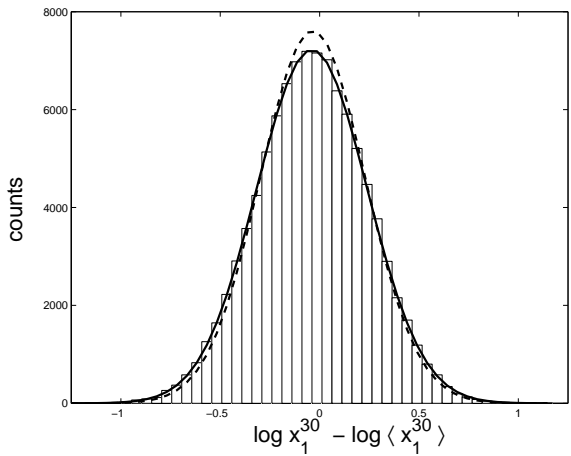}
     \caption{Approximately log-normal distribution
     of the system state}
     \label{f:hist}
\end{figure}
In the particular case of small noise and simple dominant $\LL$,
the distribution of the elements of a multivariate system is very
close to log-normal. This is shown in the simulation of figure
\ref{f:hist} which presents a histogram of the log of 100,000
instances of $x^{30}_1$ in a positive system whose $\A$ is given
in (\ref{exA}). The data were normalized by the expected value
$\lb \x^{30}_1 \rb$. The solid line is a Matlab normal fit with
$\mu = -.0389$ and $\sigma = 0.2725$. The dashed line is the
prediction of \S \ref{s:perturbationLyap} and has $\mu = -0.348$
and $\sigma = 0.2639$.

\subsection{Relevant properties of $\A$}\label{s:Aproperties}

\subsubsection{Simple dominant eigenvalue} In this paper we only
consider systems with simple, dominant $\LL$. Geometrically, the
effect of $\A$ repeatedly acting on a vector is to bring that
vector into the direction of $\uu$ and to multiply its length
repeatedly by $\LL$. The behavior of unperturbed multivariate
systems with a simple, dominant $\LL$ is thus equivalent to scalar
systems in the asymptotic limit.

The requirement that $\LL$ be simple and dominant is met in all
nonnegative systems of interest (appendix \ref{s:primitive}), so
our treatment of nonnegative systems is comprehensive. Although
many arbitrary systems meet this condition as well, some do not
and we do not attempt to treat these cases. We also neglect
systems with defective (non-diagonalizable) $\A$, which form a set
of measure 0, because the nonzero elements of $\A$ are impossible
to determine exactly in most applications.

\subsubsection{Condition of $\mathbf{\LL}$} The effect of noise on
a multivariate system, from a geometric perspective, is to perturb
both the direction and length of the vector $\x$. Noise as a small
perturbation means that a given noise matrix does not swing the
$\x$ far from the direction of $\uu$ or multiply $|\x|$ by a
factor far from $\LL$. In this regime, the dynamics are well
approximated by the dynamics of a perturbed scalar system.

The regime of small noise, for the multivariate systems, is
determined not only by the size of the noise elements but also by
the sensitivity of the system to perturbation. There exist
matrices whose eigenvalues and eigenvectors are violently affected
by even a small perturbation to the matrix
elements\cite{Golubbook,Edelman}. For a perturbation treatment, we
need to know how much the dominant eigenvalue $\LL$ and its
eigenvector $\uu$ of the system are perturbed by a given level of
noise.

The response of $\LL$ to noise is characterized by a quantity
$\kappa(\LL)$ called the \textit{condition} of $\LL$. When
$\kappa(\LL)$ is large $\LL$ is said to be ill-conditioned,
meaning that its response to a system perturbation is large with
respect to the perturbation. Even a small noise causes moment
divergence in systems with an ill-conditioned $\LL$. Conversely,
when $\kappa(\LL) = 1$, $\LL$ is said to be perfectly conditioned;
its response to a system perturbation is the smallest possible and
is on the order of the size of the perturbation. In systems with a
well-conditioned $\LL$, the perturbation approximation is
applicable to relatively large noises.

The change in $\LL$ due to a small noise matrix $\B$ (small in the
sense that \newline $|\B| = \delta \ll 1$) is given by
\begin{equation}\label{rawLLcond}
  \delta\LL \approx \vv \cdot (\B \uu)
\end{equation}
to first order in $\delta$. Taking norms, we obtain the expression
for the condition of $\LL$ in the case of normalized $\uu$:
\begin{equation}\label{LLcond}
  \kappa(\LL) = |\vv|.
\end{equation}
It is clear that $\kappa \geq 1$ by the Schwartz inequality. The
sensitivity of $\uu$ to noise may also be calculated to first
order\cite{Golubbook} and depends on the condition of $\LL$. It
also depends on the gaps $\LL - \lambda_i$ between the dominant
eigenvalue and the others, and is therefore related to the
accuracy of the approximation
\begin{equation}\label{approx}
  \A^p \approx \LL^p \uu \vv^T
\end{equation}
obtained by neglecting all $\lambda_i^p$ compared to $\LL^p$. In
the limit that $\lambda_2 \rightarrow 0$, \ref{approx} is exact
for all $p$ and the sensitivity of $\uu$ is minimized.

The level of noise which qualifies as a small perturbation must
therefore depend on $\kappa(\LL)$ and $\LL - |\lambda_2|$, which
it does, as we will show. The effect of a given level of noise is
the smallest in \textit{well-behaved} systems with a well-behaved
$\LL$ and small eigenvalue gap.

\begin{table}[htb]
\footnotesize \centering
\begin{tabular}{|c|c|c|}\hline
 Matrix type & Eigenvalue gap & Condition $\kappa(\LL)$ \\
 &  $\LL - |\lambda_2|$  & of $\LL$ \\ \hline \hline
 $\A = a\G$ & 0 & 1\\ \hline
 small $\sigma_A^2$ & small & close to 1 \\ \hline
 normal $\A$ & ? & 1 \\ \hline
 large $\sigma_A^2$ & possibly small & possibly large \\ \hline
\end{tabular}
 \caption{Properties of some types of $\A$. Recall that $\G$
is the matrix of all 1's (mean value approximation).} \label{t:A}
\end{table}
It is difficult to generally characterize the condition of $\LL$
and the eigenvalue gap in terms of more physical properties of the
matrix $\A$. What we can say is summarized in Table \ref{t:A}.

Systems close to the mean value approximation (recall $\A \approx
a\G$ where $\G$ is a matrix of 1's in the mean value
approximation; \S \ref{s:notation}) are sure to be well-behaved;
however, some systems far from the mean value approximation are
also well-behaved, as shown in figure \ref{f:vgap} below as well
as figures \ref{f:gap} and \ref{f:hen}.

The correlation between eigenvalue gap and condition number of
$\LL$ are demonstrated in the scatter plots of figure \ref{f:vgap}
which show the eigenvalue gap versus $\kappa(\LL)$ for 10000
randomly generated $5 \times 5$ matrices. The matrices were
generated from a normal distribution (top left), uniform
distribution (top right), uniform distribution with probability
1/2 and 0 with probability 1/2 (bottom left) and uniform
distribution with mean 0.2 and variance 0.02 (bottom right). For
the arbitrary matrices, only those with real $\LL$ were accepted
and the entries were normalized so that $\LL = 1$. Note the
difference in the regions plotted.
\begin{figure*}[hbt]
     \centering
     \includegraphics[width=3in]{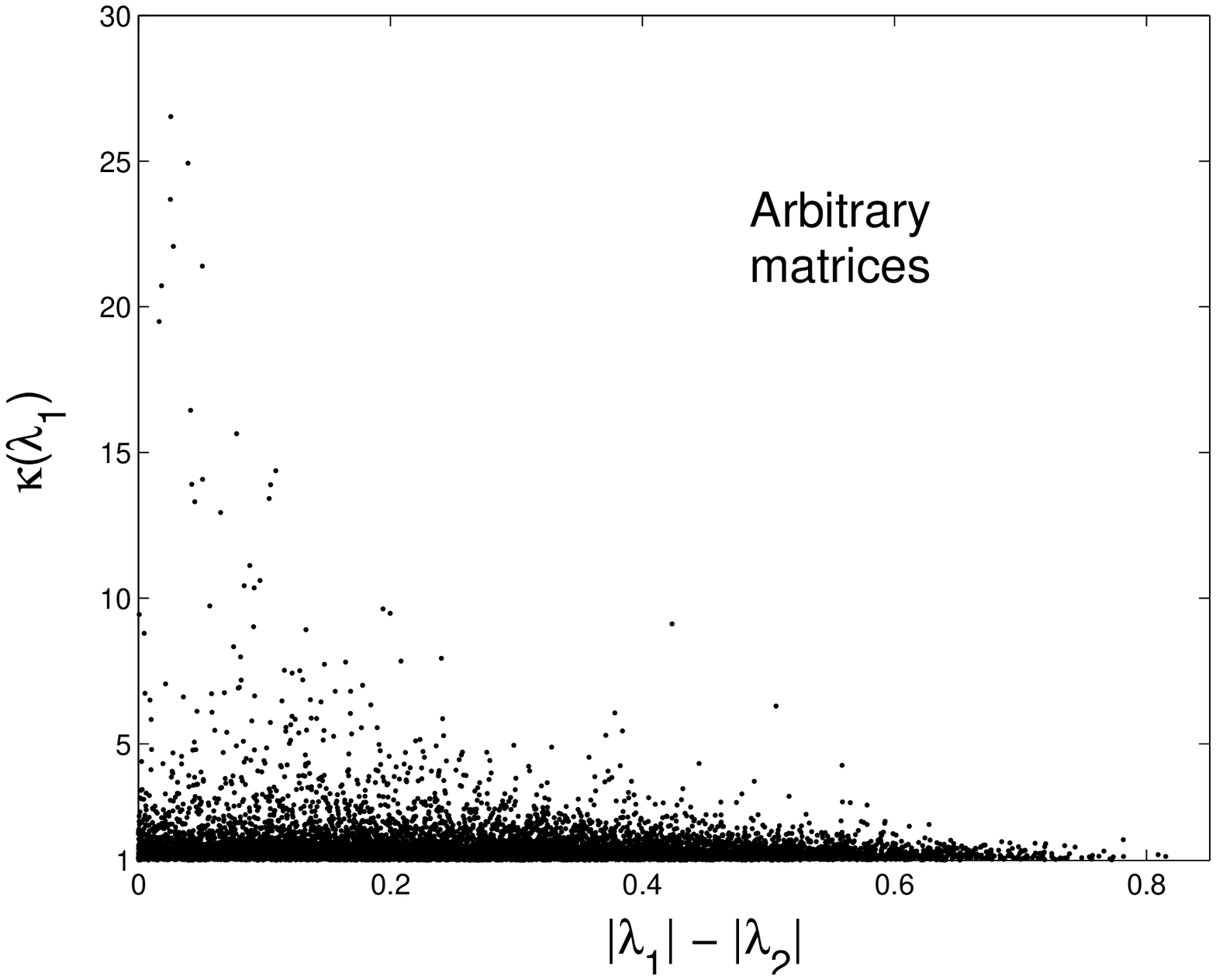}
      \includegraphics[width=3in]{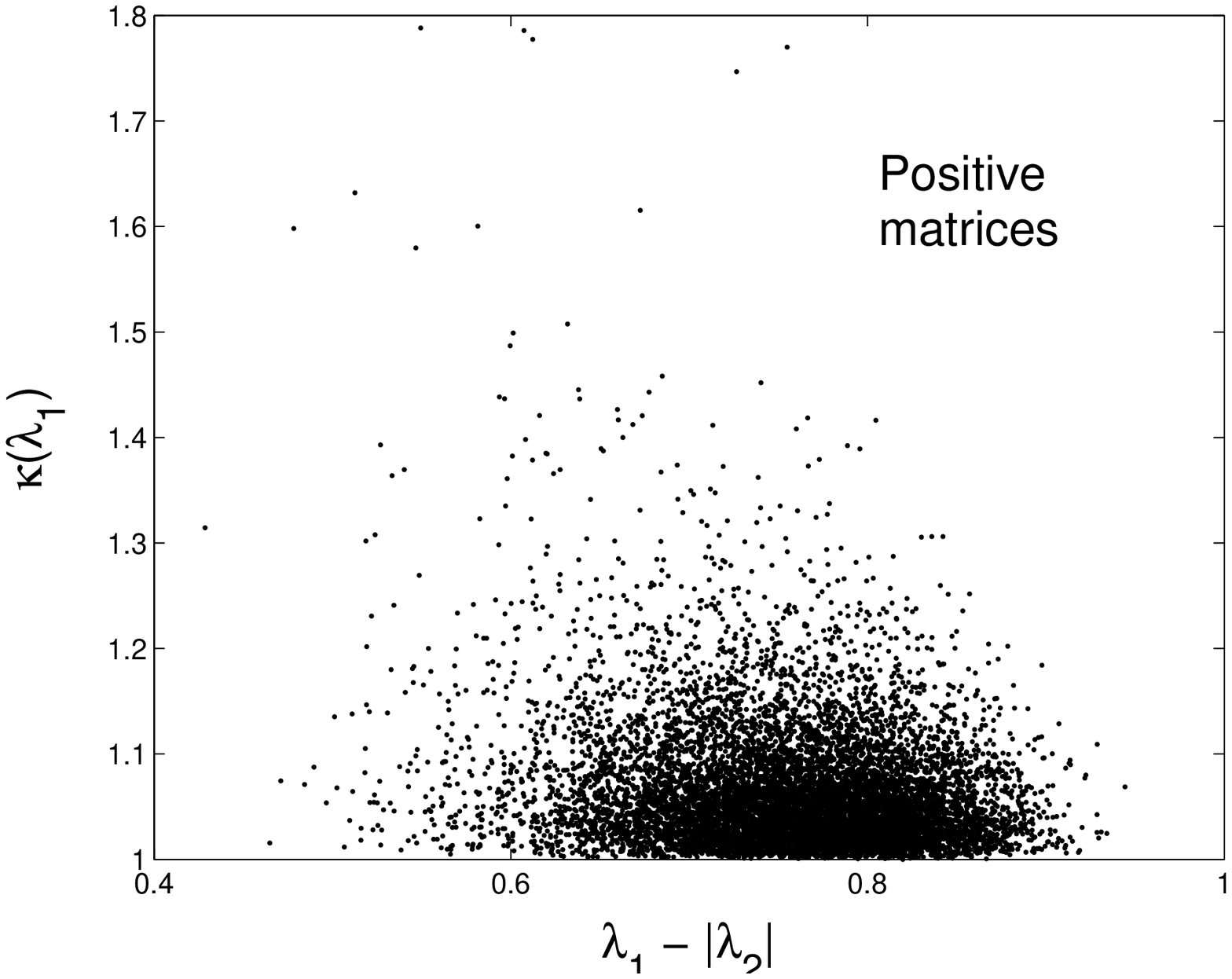} \\
     \vspace{0.1in}
\includegraphics[width=3in]{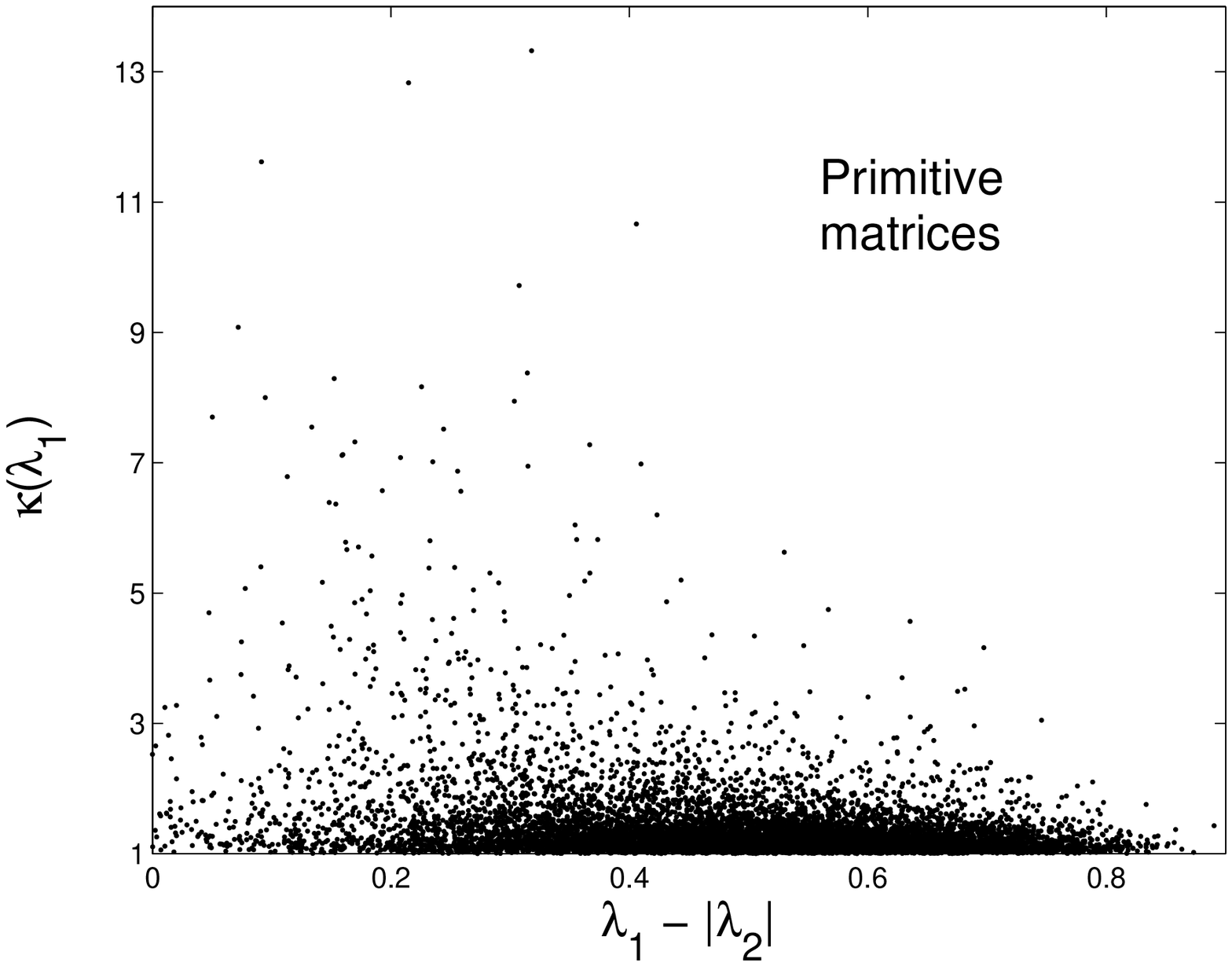}
\includegraphics[width=3in]{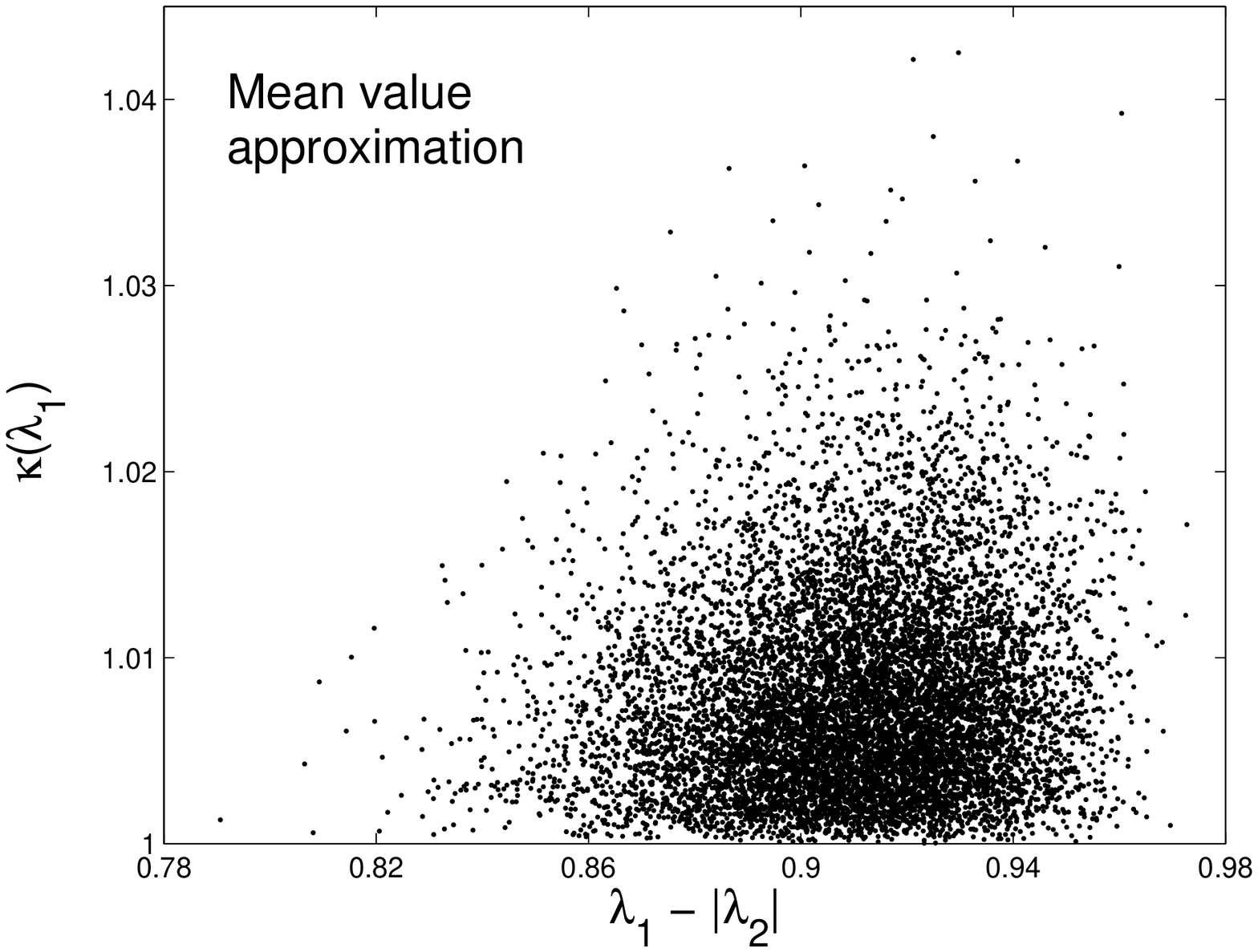}
     \caption{Eigenvalue gap versus condition of $\LL$
     in randomly generated matrices.}
    \label{f:vgap}
\end{figure*}
As shown in the figure, the likelihood that a given system is
well-behaved is larger for nonnegative matrices than for arbitrary
matrices, and larger still for positive matrices. See appendix
\ref{s:conditionetc} for further discussion.

\subsubsection{Limits on average element size} Finally we note that
in the case of nonnegative matrices, it is impossible to have a
small $\LL$ if the elements of $\A$ are too large.  Many quite
accurate bounds on the largest eigenvalue of nonnegative matrices
exist (see \cite{Minc} for a list); a relatively inaccurate but
analytically tractable bound is the row sum bound,
$\min_i(\sum_jA_{ij}) \leq \LL \leq \max_i(\sum_jA_{ij})$. This
estimate implies that on average we need to take
\begin{equation}\label{mulim}
  a < 1/n
\end{equation}
to keep $\LL < 1$ and ensure that the system converges in mean.
This is exactly the asymptotic $n$ result of \cite{Juhasz}, and
the result we would obtain in the mean value approximation $\A
\approx a\G$.

\subsection{Types of noise for multivariate systems}
\label{s:noisetypes} For multivariate systems many different forms
of noise are possible, distinguished by whether the elements are
correlated and how large their relative variances are. In this
paper we consider five cases which are analytically tractable and
have some relevance to physical systems. The correlation rules for
these cases are shown in table \ref{t:correlationrules} which
provides a summary.

For the correlation we consider three cases. \textit{Uncorrelated}
noise means that the elements of the noise matrix vary
independently. \textit{Totally correlated} noise means that all
the noise elements vary in the same way at each time step. For
symmetric systems, we consider \textit{symmetrically correlated}
noise.

For the variance we consider two possibilities. For
\textit{homogeneous} noise, the variance of every element is
identical and equal to $b^2$. For \textit{proportional} noise, the
standard deviation of $B_{ij}$ is proportional to $A_{ij}$ by some
factor $q$ which we will take to be less than 1.
\begin{table*}[htb]
\footnotesize \centering
\begin{tabular}{|l|l|}
\hline
Noise type & Correlation rule \\
\hline Uncorrelated homogeneous (UH) & $\lb B_{ij}B_{i'j'} \rb =
b^2\delta_{ii'}\delta_{jj'}$ \\
Symmetrically correlated homogeneous (SH) & $\lb B_{ij}B_{i'j'}
\rb =
b^2(\delta_{ii'}\delta_{jj'} + \delta_{ij'}\delta_{i'j})$ \\
Totally correlated (T) & $\lb B_{ij}B_{i'j'} \rb = b^2$ \\
Uncorrelated proportional (UP) & $\lb B_{ij}B_{i'j'} \rb =
q^2(A_{ij})^2\delta_{ii'}\delta_{jj'}$ \\
Symmetrically correlated proportional (SP) & $\lb B_{ij}B_{i'j'}
\rb =
q^2(A_{ij})^2(\delta_{ii'}\delta_{jj'} + \delta_{ij'}\delta_{i'j})$ \\
\hline
\end{tabular}
\caption{Correlation rules for types of noise considered in this
paper} \label{t:correlationrules}
\end{table*}

\section{Small noise as a perturbation} \label{s:perturbation}
In this section we determine approximate expressions for the
moment Lyapunov exponents for multivariate systems subject to
small noise using a perturbation treatment. We examine the
dependence of the Lyapunov exponents on system properties, and
discuss the accuracy of the approximation.

First let us reexpress the matrix product in (\ref{main2}):
\begin{eqnarray}
\x^t &=& \left[ \prod_{\tau=1}^t(\A + \B^\tau) \right]\x^0 \label{prod} \\
&=&\left[
\sum_{\Y^t=\A,\B^t}\sum_{\Y^{t-1}=\A,\B^{t-1}}\sum_{\Y^1=\A,\B^1}\Y^t\Y^{t-1}\cdots
\Y^1 \right]\x^0 \nonumber \\
&=& \sum_{\textnormal{each } \Y^\tau=\A,\B^\tau}\Y^t\Y^{t-1}\cdots
\Y^1 \x^0 \label{Ys}
\end{eqnarray}
meaning that each $\Y^\tau$ in the sum can be either $\A$ or
$\B^\tau$, for $\tau = 1\ldots t$. There are $2^t$ terms in the
sum; each term is a vector.

\subsection{Perturbation expansion} \label{s:expansion}
The perturbation expansion consists of considering only terms in
(\ref{Ys}) which have very few $\B$'s.  For small noise, these
terms make the only important contribution to the sum. Let us
assume that this is so without justification, even before we
define small noise.

The reason that this strategy simplifies the calculation is as
follows. Consider the evolution in time of the length and
direction of a single term of (\ref{Ys}) with few $\B$'s. In the
asymptotic limit, a typical term with few $\B$'s has long strings
of consecutive $\A$'s broken by single occurrences of $\B$'s. As
far as the direction of such a term, the long strings of $\A$'s
act to bring it parallel to $\uu$ as previously mentioned (see
(\ref{approx})). When a $\B^\tau$ acts on the term, the term lies
almost parallel to $\uu$; even though the noise causes the term to
point away from $\uu$, the next string of $\A$'s brings it back to
the direction of $\uu$ before another noise term occurs. The
action of $\B^\tau$ is thus independent of $\tau$. As to the
length, a string of $p$ $\A$'s simply multiplies the term length
by $\LL^p$; and the $\B$'s multiply the length by some stationary
random variable.

In a term with few $\B$'s, therefore, the position of the matrices
in the sum (\ref{Ys}) is unrelated to their net effect on the
term. Thus the matrices in the sum can be replaced by scalars, and
the matrix product (\ref{prod}) becomes a product of scalars. To
illustrate this, consider a typical term for $t=10$ with a
$\B^\tau$ only in the $\tau = 6$ spot:
\begin{eqnarray*}
\A\A\A\A\B^6\A\A\A\A\A\x^0 &\approx& (\LL^4 \uu\vv^T)(\B^6)(\LL^5\uu\vv^T) \x^0 \\
&\approx& \LL^{10} \varepsilon_6 \uu (\vv \cdot \x^0)
\end{eqnarray*}
where we define the random variable
\begin{equation}\label{eps}
  \varepsilon_\tau = \frac{\vv^T\B^\tau\uu}{\LL}.
\end{equation}
In general, a term of the sum (\ref{Ys})  that has long strings of
$\A$'s and $m$ isolated $\B$'s $\{\B^{\tau_1},\ldots,
\B^{\tau_m}\}$ points in the direction of $\uu$ and has length
$\LL^t (\varepsilon_{\tau_1}\cdot \ldots \cdot
\varepsilon_{\tau_m}) \uu (\vv \cdot \x^0)$. Such terms dominate
the sum (\ref{Ys}) (see \S \ref{s:accuracy}) and so the system
state is given approximately by
\begin{equation}\label{Pt}
  \x^t \approx \uu (\vv \cdot \x^0)\LL^t \prod_{\tau = 1}^t (1 + \varepsilon_\tau).
\end{equation}
The random variables $\varepsilon_\tau$ are i.i.d. and satisfy
$\lb \varepsilon_\tau \rb = 0$; the moments depend on the form of
the noise. Notice that the numerator of $\varepsilon_\tau$ is
exactly equal to the first order change in $\LL$ due to a small
perturbation to $\A$ (\ref{rawLLcond}) and thus closely related to
the condition $\kappa(\LL)$ (\ref{LLcond}). The eigenvalue gap and
thus the sensitivity of $\uu$ is implicitly involved in this
expression from the application of (\ref{approx}).

\subsection{Criterion for small noise} \label{s:smallnoise}
The simplest small noise criterion is
\[
  \lb \varepsilon_\tau^p \rb \ll 1
\]
for all $p$. This is a rather complicated condition since the
calculation of all the moments can be difficult for some forms of
noise. Instead we choose a more restrictive (triple) condition,
\begin{eqnarray}
P(|\varepsilon_\tau| > 1) &=& 0  \nonumber\\
\lb \varepsilon_\tau^p \rb &\sim& \varepsilon^{p'}, \ \ p' \leq p \label{small} \\
  \varepsilon^2 &\ll& 1 \label{small2}
\end{eqnarray}
Note that this requirement is not trivial as in the scalar case
because the condition of $\LL$ can be large. Less restrictive
conditions are possible but this will enable us to better
understand the dynamics by taking logs and expanding in a power
series in $\varepsilon$.

\subsection{Moment evolution} \label{s:perturbationLyap}
Using the perturbation expansion of section \ref{s:perturbation},
we may now present approximate expressions for the moments of a
multivariate stochastic system. We do so by calculating the
approximate Lyapunov exponents, proceeding from (\ref{Pt}) exactly
as in the scalar case of \S \ref{s:scalarapprox} with
$\varepsilon_\tau$ playing the role of $b_\tau$ and $\LL$ the
multidimensional analog of $a$.

We thus find
\begin{eqnarray}
L_p &\approx& p\log \LL - p(p-1) \lb \ln(1 + \varepsilon_\tau) \rb  \nonumber \\
&\approx& L^0_p + p(p-1)\frac{\varepsilon^2}{2} + O(\varepsilon^3)
\label{L}
\end{eqnarray}
where
\[
L^0_p = p \log \LL
\]
is the Lyapunov exponent for the unperturbed system and the error
is $O(\varepsilon^4)$ if the noise is symmetric. The system
moments are
\begin{eqnarray}
\lb |\x^t|^p \rb &\approx& |\vv \cdot \x^0|^p e^{tL_p} \nonumber \\
&\approx& |\lb \x^t \rb|^p e^{t\varepsilon^2\frac{p(p-1)}{2} +
O(\varepsilon^3)}. \label{moments}
\end{eqnarray}
In particular,
\begin{equation}\label{L2}
  L_2 \approx L^0_2 + \varepsilon^2
\end{equation}
to second degree in $\varepsilon$, and
\begin{equation}\label{E2}
  \lb |\x^t|^2 \rb \approx |\lb \x^t \rb|^2 e^{t\varepsilon^2}.
\end{equation}
Notice that to this level of approximation, first moment (norm)
convergence is not distinguishable from convergence in mean.

\begin{table}[hbt]
\footnotesize \centering
\begin{tabular}{|c|l|}
\hline Noise Type & $\varepsilon^2$ \\
\hline
UH & $\frac{v^2b^2}{\LL^2}$ \\
SH & $\frac{2b^2}{\LL^2}$ \\
T & $\frac{b^2}{\LL^2} (\Sigma_iv_i)^2
(\Sigma_iu_i)^2$ \\
UP & $\frac{fq^2}{\LL^2} \sum_{ij}v^2_i(A_{ij})^2u_j^2$ \\
SP & $\frac{fq^2}{\LL^2}
\sum_{ij}(A_{ij})^2[v_i^2u_j^2 + v_iv_ju_iu_j] $ \\
\hline
\end{tabular}
\caption{Values of $\varepsilon^2$ for the types of noise defined
in \S  \ref{s:noisetypes}} \label{t:varepsilonnoises}
\end{table}
To proceed beyond these expressions we must evaluate
$\varepsilon^2 = \lb (\frac{\vv^T\B\uu}{\LL})^2 \rb$, which we
cannot do without specifying the form of the noise. The values of
$\varepsilon^2$ for the noises described in \S \ref{s:noisetypes}
are easily calculated and presented in table
\ref{t:varepsilonnoises}. Here UP is uncorrelated proportional
noise, UH is uncorrelated homogeneous, SP is symmetrically
correlated proportional, SH is symmetrically correlated
homogeneous, and T is and totally correlated. In the proportional
noise, $f$ is a factor which depends on the distribution chosen;
for example, $f = 1$ for normal noise and $f = 1/3$ for uniform
noise. In the case of symmetrically correlated noise, a symmetric
$\A$ is assumed.

\begin{figure*}[hbt]
     \centering
      \includegraphics[width=3in]{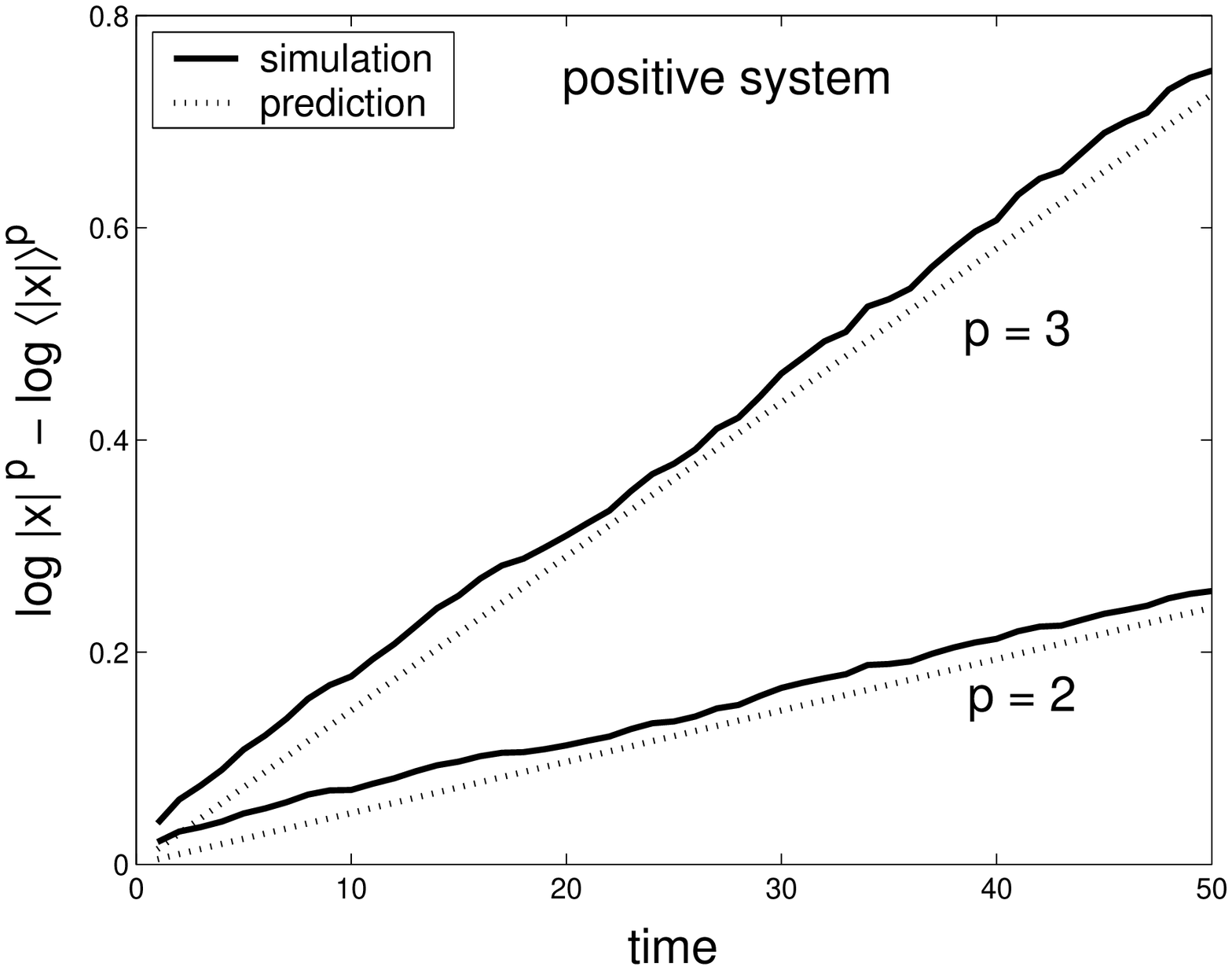}
\includegraphics[width=3in]{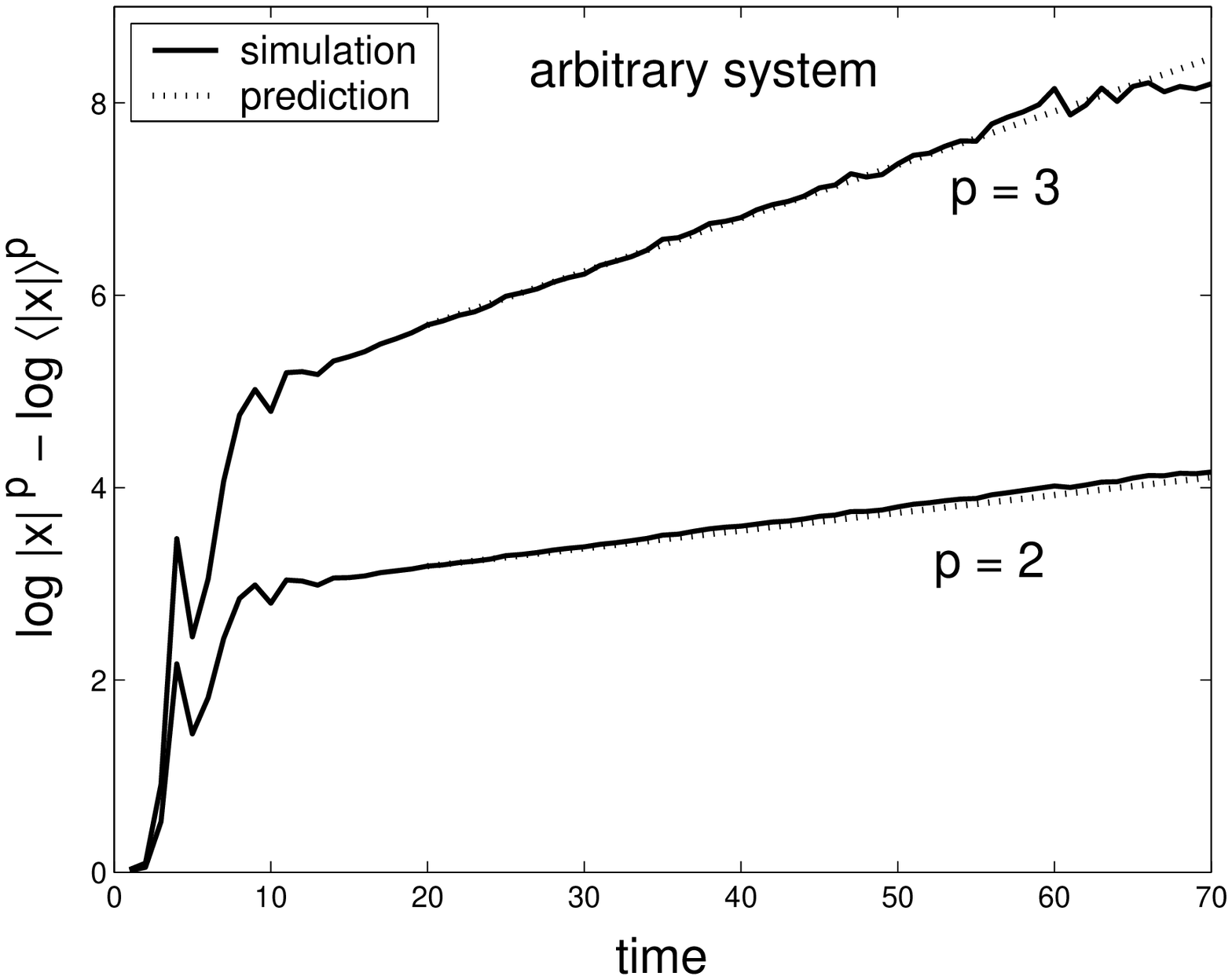}
     \caption{Moment evolution in two systems}
    \label{f:L}
\end{figure*}
The accuracy of the above approximations for the moment evolution
is demonstrated in figure \ref{f:L}. In this figure, the log of
the 2nd and 3rd moments of two randomly generated systems are
shown. The plots are normalized by the expected value (unperturbed
value) of the system and show only the noise part. The solid line
is the average over 10,000 runs of the simulation, and the dotted
line shows the analytic prediction of (\ref{moments}). At left is
the positive system of (\ref{exA}) subjected to uniform UP noise
with $q = 0.5$. Note that the asymptotic limit is reached almost
immediately in this system. At right is an arbitrary system with
simple dominant $\LL$ subjected to normal UH noise with $b = 0.1$.
This system has large transient behavior before it settles in to
its asymptotic limit around $t = 20$. The analytic prediction,
which cannot account for the transient, has been artificially
placed to demonstrate the asymptotic accuracy of the slope.

\subsection{Dependence on system size} \label{s:ndependence}
We can now explore the $n$ dependence of the moments. Because the
small noise case is important for applications, we present a
detailed discussion of the size dependence based only on the
expressions developed thus far. A different discussion of the $n$
dependence for larger noises is presented in \S
\ref{s:crit_n_dependence}. For simplicity, we consider only the
second moment in this section.

As we show, independently varying noises ``interfere'' with each
other and diminish the effect of the noise, compared to the
unperturbed system. Thus, the effect of the noise decreases as $n$
increases in the case of uncorrelated noise. There is no $n$
dependence to second order, however, in the case of totally
correlated noise. Symmetrically correlated noise provides an
intermediate case.

We also show that for noise proportional to the system elements,
as the system deviates from the mean value approximation and in
particular becomes closer to diagonal, the destructive
interference is decreased and the noise has a greater effect.

\subsubsection{Mean value approximation}
As a first simplification, we consider the mean value
approximation where $\A \approx a\G$. In this case, $v_i \approx
u_i \approx 1/\sqrt{n}$ for all $i$, and $\LL \approx na$. We
consider homogeneous noise (all the noise elements have the same
variance, which is almost equivalent to proportional noise in the
mean value approximation) with variance $b^2 = q^2a^2$, $q<1$.

\begin{table}
\footnotesize \centering
\begin{tabular}{|l|l|}
\hline
Noise Type & $L_2$ \\
\hline
UH & $L_2 \approx L^0_2 + \frac{q^2}{n^2}$  \\
SH:  & $L_2 \approx L^0_2 +
\frac{q^2}{n^2/2}$ \\
T:  & $L_2 \approx L^0_2 + q^2$ \\
\hline
\end{tabular}
\caption{Approximate value of $L_2$ for three types of homogeneous
noise} \label{t:sizedependence}
\end{table}
Using (\ref{L2}) and table \ref{t:varepsilonnoises}, the values of
$L_2$ for three types of homogeneous noise are easily computed in
the mean value approximation and are shown in table
\ref{t:sizedependence}. These expressions are to be compared to
the scalar case $L_2 \approx L^0_2 + q^2$ (equation \ref{L2scal}).

Note in particular how the noise effect (the $q^2$ term) is
divided by a factor related to the number of independent elements
of the noise. This destructive interference is not surprising when
we consider why multiplicative noise processes generate the
anomalously large events which make up the heavy tail of the
log-normal distribution. The anomalous events result from a long
sequence of large, positive noises \cite{02thing}. When there are
$n^2$ independent noises per time step, as opposed to 1, anomalous
events are rarer. However, when all the elements of noise vary
identically, the effect of the noise is the same in scalar and
multidimensional systems. Note that symmetric noise provides an
intermediate calculable case; there are $n(n+1)/2$ independent
components in a symmetric noise.

\begin{figure*}[hbt]
     \centering
      \includegraphics[width=3in]{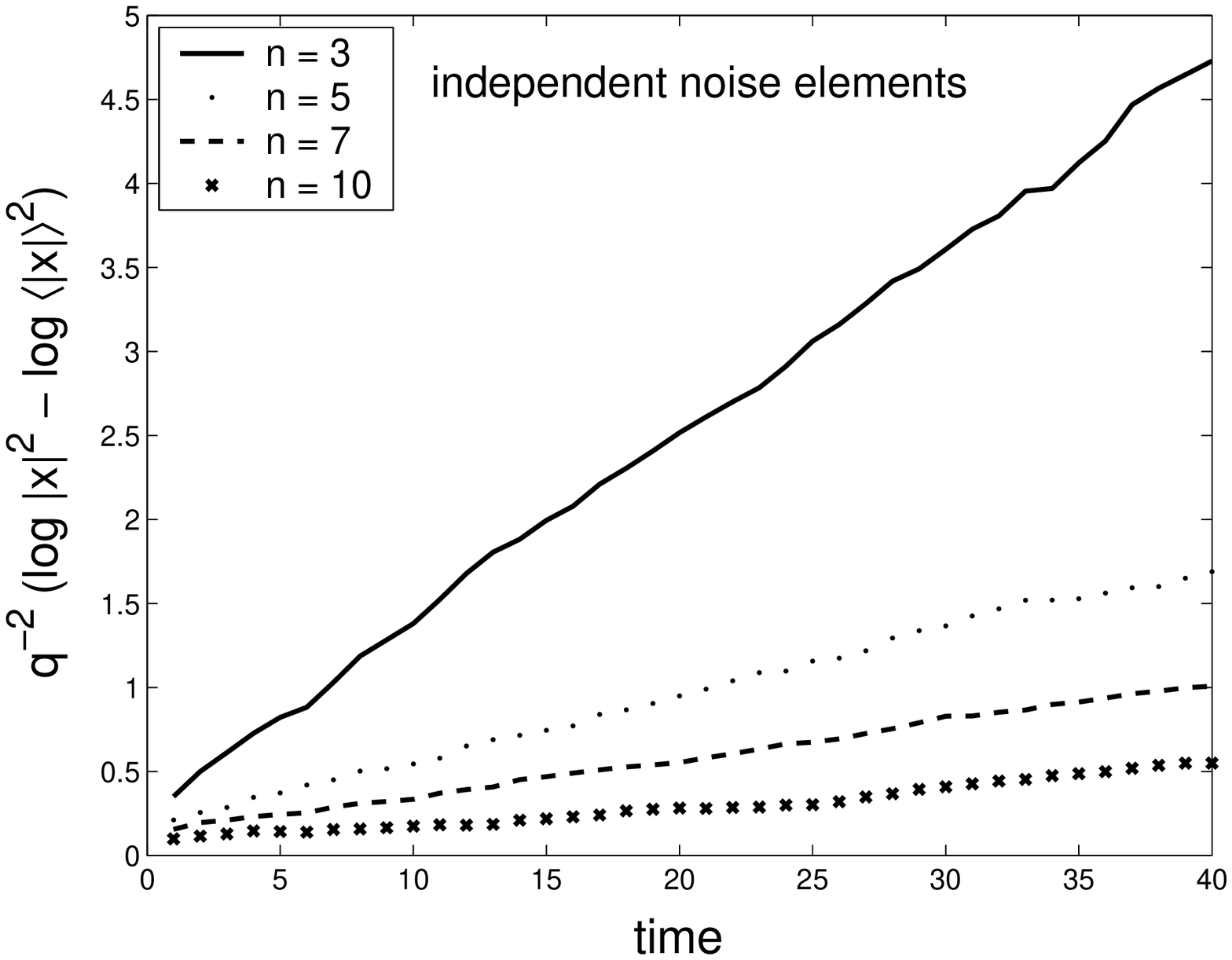}
     \includegraphics[width=3in]{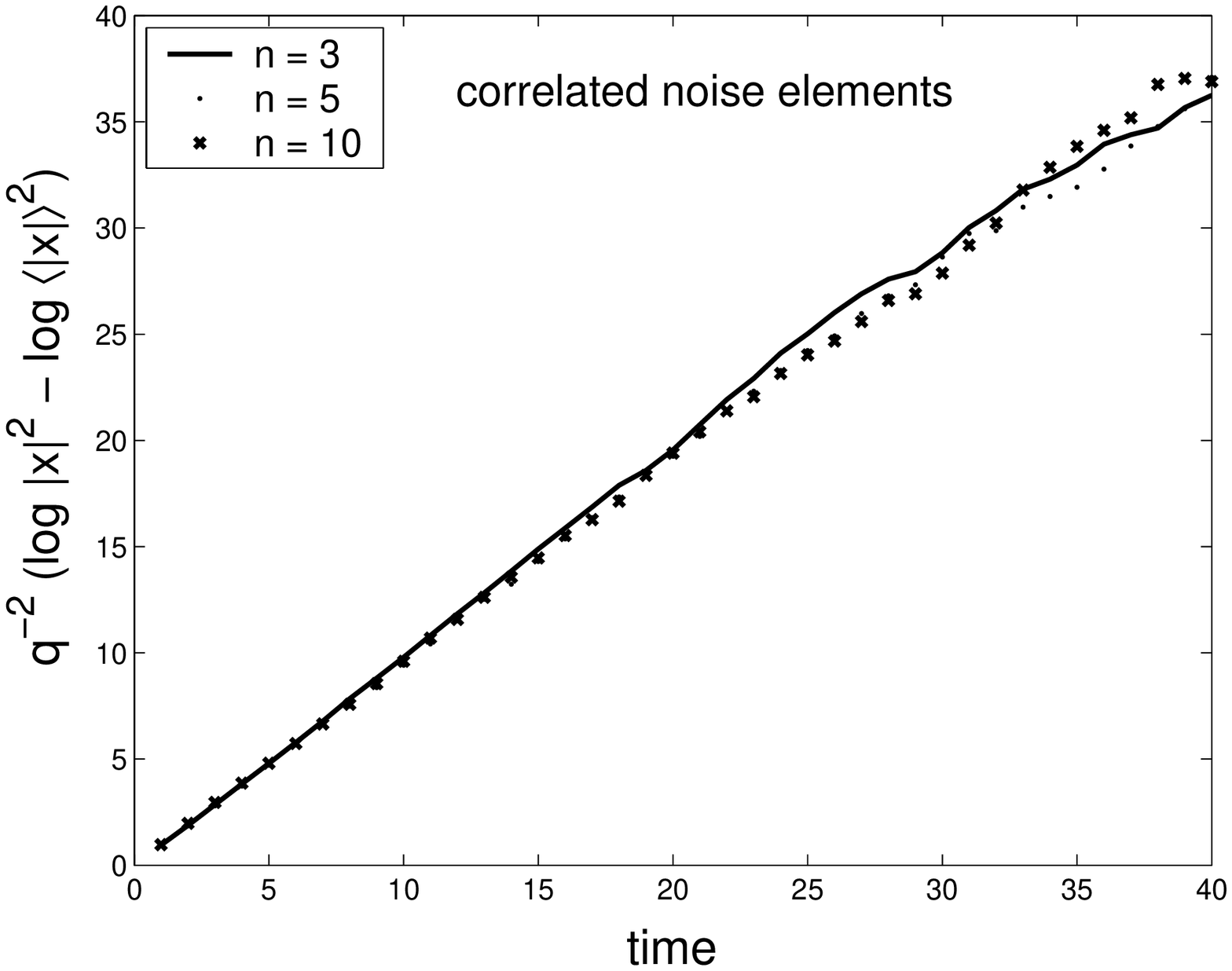}
     \caption{Dependence of the noise
     effect on the number of independent noise elements.}
    \label{f:sizedependence}
\end{figure*}
The dependence of the noise effect on the number of independent
elements of the noise is demonstrated in the simulation of figure
\ref{f:sizedependence}.  By table \ref{t:sizedependence} and
equation (\ref{E2})
\[
\lb |\x|^2 \rb/|\lb \x \rb|^2 = e^{tfq^2} \]  in the asymptotic
limit, where $f = 1/n^2$ for UH noise (all noise elements
independent) and $f = 1$ for T noise (all noise elements
correlated). Accordingly we plot \newline $q^{-2}(\log |\x|^2 -
\log |\lb \x \rb|^2)$, averaged over 100,000 runs, for various
values of $n$. On the left is the plot for UH noise, where the
lines should have slope $1/n^2$; on the right is the plot for T
noise where the slope should be 1 for any $n$. The agreement is
excellent. The $\A$'s for these systems were randomly generated
from uniform distributions with small variance. The initial state
was a vector of 1's, and $q = 1/4$. Because we are in the mean
value approximation, $\lambda_2$ is very small for these systems
and the asymptotic limit for $t$ begins almost immediately.

\subsubsection{Deviation from the mean value approximation}
We examine the effect of a deviation from the mean value
approximation on $L_2$ in the case of uncorrelated proportional
noise. The result is that the noise effect roughly increases the
larger the deviation, as the $1/n^2$ damping caused by the
independent noise elements is mitigated. This is because we have
assumed that the typical size of a noise element is proportional
to the corresponding element of $\A$ (UP noise), so that small
entries of $\A$ contribute little to the interference effect of
the independent noises.

In this subsection we assume that the approximation (\ref{approx})
is accurate for $p = 2$ so that $A_{ij}^2 \approx
\LL^2u_i^2v_j^2$. Recall that (\ref{approx}) is generally more
accurate the closer $\A$ is to the mean value approximation, but
it can be accurate even if the variance of the $A_{ij}$ is large,
as discussed above.

With the above approximation we have
\begin{equation}\label{wL2}
  L_2 \approx \ln\LL^2 + q^2w^4
\end{equation}
where we define
\[
w^2 = \sum_{i} v^2_iu^2_i.
\]
Comparing (\ref{wL2}) to the mean value case, we see that the
$1/n$ factor is replaced by $w^2$. This quantity satisfies
\[
1/n \leq w^2 \leq 1
\]
since $\vv \cdot \uu = 1$. The lower bound is achieved in the mean
value case; the upper bound is achieved when $\A$ is diagonal.
$w^2$ is thus a rough measure of the deviation of the $A_{ij}$
from the mean value approximation; it generally increases as the
variance of the $A_{ij}$ increases.

Larger values of $w^2$ for deviations from the mean value
approximation thus mean a larger noise effect. This effect is
demonstrated in figure \ref{f:wsizedependence}, which is analogous
to figure \ref{f:sizedependence} and compares the noise parts of
the second moment $q^{-2}(\log |\x|^2 - \log |\lb \x \rb|^2)$ of a
mean value approximation (MVA) system and of a system with the
same $\LL$ and $n=5$ but whose elements have a much larger
standard deviation. UP noise was considered, so the slope should
be $w^4$ by (\ref{wL2}). The dashed line system has $w^2 =
0.2888$, $w^4 = 0.0834$ while the MVA system has $w^2 = 0.2002
\approx 1/n$ and $w^4 = 0.04$; a rough linear interpolation fit to
the data shows an asymptotic slope 0.095 for the dashed system and
0.0385 for the MVA system. Note how the dashed system does not
immediately reach its asymptotic limit; it has $|\lambda_2| \sim
\LL/2$ as opposed to $|\lambda_2| \ll 1$ for the MVA system.
\begin{figure}[hbt]
     \centering
      \includegraphics[width=3in]{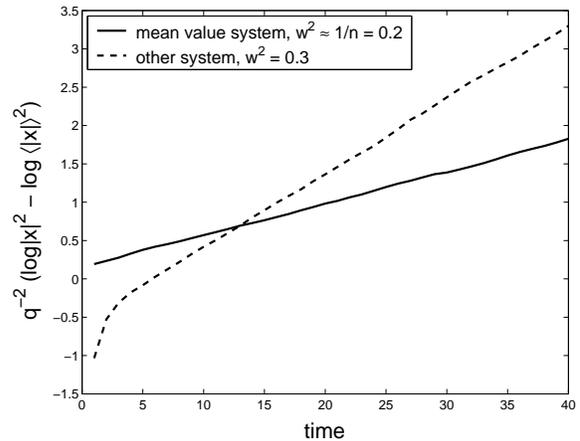}
     \caption{Larger noise effect of a deviation from the mean value approximation}
    \label{f:wsizedependence}
\end{figure}

\subsubsection{Large $n$ limit and homogeneous noise}
When the noise is homogeneous we can apply results on spectral
theory of matrices to study the $n$ dependence in the large $n$
limit without appealing to a mean value approximation. In the case
of a symmetric matrix with entries drawn from a distribution with
mean $a$ and variance $\sigma^2_a$
\[
\LL = na + \sigma^2_a/a
\]
on average \cite{FurediKomlos}. For an arbitrary (asymmetric)
matrix \cite{Juhasz}
\[
\LL \sim na,
\]
which is really just the mean value approximation.

\begin{table}[hbt]
\footnotesize \centering
\begin{tabular}{|ll|}
\hline Arbitrary system & $L_2 \sim L^0_2 + \frac{b^2/a^2}{n^2}$  \\
\hline Symmetric system  & $L_2 \sim L^0_2 + \frac{b^2/a^2}{(n^2 +
2n\sigma_a^2/a^2)/2}$ \\ \hline
\end{tabular}
\caption{Large $n$ dependence of arbitrary and symmetric systems
with homogeneous noise.} \label{t:largenndep}
\end{table}
We thus obtain table \ref{t:largenndep} for the $n$ dependence.
Recall the notation $b^2$ for the variance of the homogeneous
noise. Note again the $n^2$ damping in the arbitrary system, and a
damping on the order of $n(n+1)/2$ in the symmetric case, in
agreement with the previous analysis. We have assumed
independently varying noise in the arbitrary system, and
symmetrically varying noise in the symmetric system.

\subsection{Approximation justification, accuracy, failure}
\label{s:accuracy} The justification for equation (\ref{Pt}) in
the small noise approximation is as follows. Expand the product
$\prod_\tau(1 + \varepsilon_\tau)$ into a sum. The typical size of
the random variable $\varepsilon_\tau$ is $\varepsilon$, and the
largest contribution to the sum comes from terms with $k_{\max}$
$\varepsilon_\tau$'s, where
\[
k_{\max} = [t\varepsilon]
\]
is the binomial expected value. The brackets denote the closest
integer. This means that the largest terms in the sum come from
terms of (\ref{Ys}) with $k_{\max}$ $\B$'s.  From symmetry
considerations it is clear that in the asymptotic limit, the
average separation $d$ between two $\B$'s in a term with
$k_{\max}$ $\B$'s is
\[
\lb d \rb = \frac{t}{1 + t\varepsilon},
\]
which is large for small $\varepsilon$ and asymptotically
independent of $t$. Furthermore, in a term with $k_{\max} \approx
t\varepsilon$ $\B's$, the separation satisfies\cite{Rice}
\[
P(d < d_0) = 1 - (1-\frac{d_0}{t})^{t\varepsilon} \rightarrow \varepsilon d_0 \\
\]
in the asymptotic limit, which is small for small $\varepsilon$
and independent of $t$. Therefore, the important terms of
(\ref{Ys}) for small $\varepsilon$ are those with a few $\B$'s
separated by long strings of $\A$'s for all $t$ \footnote{Note
that simulation of divergent moments in a convergent system for
large $t$ may not seem accurate, because as $t$ increases the
probability of an anomalous event becomes very small. A very large
sample space is necessary to obtain an accurate simulation for
large $t$; see the discussion in \cite{Redner}}.

However, this analysis does not tell the entire story. The
accuracy of the perturbation approximation is in fact much higher
than one would expect from the above calculation. To understand
this, consider a term of the sum (\ref{Ys}) with many $\B$'s. This
term's direction is impossible to determine in general because
each noise matrix transforms it arbitrarily. There are many such
terms and they are all affected by a different set of noise
matrices. Their directions are thus widely distributed in
$\mathbf{R}^n$ and mostly cancel out in the sum.

When $\varepsilon$ is not small terms in the sum (\ref{Ys}) with
many $\B$'s become important. This causes the perturbation
approximation to be inaccurate for two different reasons. First,
when $\B$'s are adjacent, the approximation of replacing $\B$ by
$\varepsilon_\tau$ is poor; second, when there are many strings of
only a few adjacent $\A$'s, both replacing $\A$ by $\LL^2$ and
$\B^\tau$ by $\varepsilon_\tau$ can be inaccurate. The relative
importance of these two inaccuracies can be different. For
example, the accuracy of the $\A$ factor is independent of $n$
while the accuracy of the $\B$ factor decreases as $n$ increases.

It is difficult to determine a cut-off where $\varepsilon$ becomes
large. The overall error may be much smaller than the error of
each term of the sum (\ref{Ys}), because the deviations of the
terms may lie in different directions and cancel out in the sum.
It is clear that the cut-off depends on how quickly $\A^p$ brings
a random vector into alignment with $\uu$, but even this is a
complicated function of the eigenvalue gap and the condition of
$\LL$ \cite{Golubbook}. To account for large $\varepsilon$ and
handle the contribution from neighboring $\B$'s accurately for
large $n$, we develop a different approximation in the next
section.

\section{Arbitrary noise using iteration approximation}
\label{s:iteration} We now present a different method, the
iteration technique, which can be applied to find an approximate
value for the second moment for small or large noise in well- or
ill-conditioned systems.

For homogeneous noise (that is, all the noise elements having the
same variance), the approximation can be extended to any level of
accuracy for any noise. Unfortunately, for other forms of noise
including proportional noise, only the first approximation is
applicable.

In addition to providing a way to treat systems where the noise
effect is not small, this technique is able to detect the explicit
$n$-dependence of the noise effect. This effect is very slight in
the small noise case, but quite important for larger noises.

The general strategy of the method is to express  $\lb
|\x^{t+1}|^2 \rb$ as a time-independent function of  $\{\lb
|\x^t|^2 \rb ,\lb |\x^{t-1}|^2 \rb,\ldots\}$ in the asymptotic
limit. A similar technique was independently developed in
\cite{Roy} for other applications.

\subsection{First approximation}
\label{s:iter_first_approx} The first approximation of this method
consists of applying the relation
\begin{equation}\label{app2}
  \sum_{ij}(A^r_{ij})^2 \approx \LL^{2r} v^2,
\end{equation}
where $\uu$ has been normalized to have length 1, for all $r$,
even $r=1$. In this approximation we can express $\lb |\x^{t+1}|^2
\rb$ as a $t$-independent function of $\lb |\x^t|^2 \rb$ alone, as
we will see. We thus define $\x_A^t = \A\x^{t-1}$ and $\x_B^t =
\B\x^{t-1}$, so that $\x^t = \x_A^t + \x^t_B$. We have
\begin{equation}\label{xt0}
  \lb |\x^t|^2  \rb = \lb |\x_A^t|^2 \rb + \lb |\x_B^t|^2 \rb;
\end{equation}
the cross term is zero in expectation because there is one power
of $\B^t$. We will establish a matrix recurrence relation
\begin{equation}\label{M1}
   \left( \begin{array}{c}
     \lb |\x_A^{t+1}|^2 \rb \\
     \lb |\x_B^{t+1}|^2 \rb \end{array} \right) \approx
      \M_1
 \left( \begin{array}{c}
     \lb |\x_A^t|^2  \rb\\
     \lb |\x_B^t|^2 \rb \end{array} \right)
\end{equation}
where the elements of $\M_1$ (subscript 1 for first approximation)
are independent of time. The asymptotic behavior of the second
moment is $\lb |\x^t|^2 \rb \sim \mu_1^t$, where $\mu_1$ is the
largest eigenvalue of $\M_1$, and
\[
L_2 \approx \ln \mu_1
\]
for the Lyapunov exponent.

Using the new notation on the recurrence relation, we have:
\begin{eqnarray}
\lb |\x^{t+1}|^2 \rb &=& \sum_{ijj'} \left\lb  (A +
B^{t+1})_{ij}(x_A^t + x_B^t)_j \right. \\
\nonumber & & \left.  (A + B^{t+1})_{ij'}(x_A^t + x_B^t)_{j'}
\right\rb
\end{eqnarray}
or
\begin{eqnarray}\label{base}
& & \lb |\x^{t+1}|^2 \rb = \lb |\A\x_A^t|^2 \rb +
 \sum_{ijj'}A_{ij}A_{ij'}\lb (x_B^t)_j(x_B^t)_{j'}
 \rb + \\
& & + \sum_{ijj'}\lb B_{ij}B_{ij'}\rb \lb (x_A^t)_j(x_A^t)_{j'}\rb
+
 \sum_{ijj'}\lb B_{ij}B_{ij'} \rb \lb (x_B^t)_j (x_B^t)_{j'} \rb \nonumber
\end{eqnarray}
because the noise is white with mean 0. This is the simplest form
we can obtain without considering particular types of noise.

\subsubsection{Homogeneous noise} Recall that homogeneous noise is a type
of noise in which all the elements of $\B$ have the same variance.
In this case, equation (\ref{base}) becomes
\begin{eqnarray}\label{first_approx_equations}
    \lb |\x^{t+1}|^2 \rb &\approx&
\LL^2 \lb |\x_A^t|^2 \rb + (\LL^2f_v/n)\lb |\x_B^t|^2 \rb + \\
\nonumber & & \!\!\!\!\! + (n^k b^2 f_u) \lb |\x_A^t|^2 \rb +
(n^kb^2) \lb |\x_B^t|^2 \rb)
\end{eqnarray}
where we introduce the notation
\begin{equation}\label{f_v}
f_v = \left\{ \begin{array}{cc} v^2, & \textnormal{UH noise} \\
(\Sigma_iv_i)^2, & \textnormal{T noise} \end{array} \right.
\end{equation}
\begin{equation}\label{f_u}
f_u = \left\{ \begin{array}{cc} 1, & \textnormal{UH noise} \\
(\Sigma_iu_i)^2, & \textnormal{T noise} \end{array} \right.
\end{equation}
as well as the factor
\begin{equation}\label{k}
  k = \left\{ \begin{array}{cc} 1, & \textnormal{UH noise} \\
  2, & \textnormal{T noise} \end{array} \right.
\end{equation}
to account for the difference between independent (UH) noise and
correlated (T) noise. Multiple steps have been skipped in
obtaining equation (\ref{first_approx_equations}), including the
use of (\ref{approx}) with $p=2$ on the first term and
(\ref{app2}) on the others.

We thus obtain
\[
\M^{UH}_1 = \left(
\begin{array}{cc}
     \LL^2 & f_v\LL^2/n\\
     nf_ub^2 & n^kb^2 \end{array} \right).
\]
and the second moment evolves as $\mu_1^t$, where the largest
eigenvalue $\mu_1$ of $\M^{UH}_1$ is
\begin{equation}\label{mu_UH}
  \mu_1 = \frac{\LL^2 + n^kb^2 + \sqrt{(\LL^2 - n^kb^2)^2 +
  4\LL^2v^2b^2}}{2}.
\end{equation}
Recalling that for UH noise $v^2b^2/\LL^2 = \varepsilon^2$, while
for T noise $b^2(\Sigma_iu_i)^2(\Sigma_iv_i)^2/\LL^2 =
\varepsilon^2$, we thus have
\begin{eqnarray}\label{iterapproxL2}
\nonumber L^{UH}_2 &\approx& \ln[\LL^2(\frac{1 + n^kb^2/\LL^2 +
\sqrt{(1 -
n^kb^2/\LL^2)^2 + 4\varepsilon^2}}{2})] \\
&\approx& L^0_2 + \varepsilon^2 + \varepsilon^4(\frac{n^k}{v^2} -
\frac{3}{2}) + O(\varepsilon^6),
\end{eqnarray}
where $L^0_2 = 2 \log \LL$, and the approximation in the second
line is valid in the limit of small $\varepsilon^2$ \textit{and}
small $nb^2/\LL^2$. The main difference between this expression
and the perturbation expansion is that we have taken into account
the effect of two neighboring $\B$'s, which produces a factor of
$n$. The $n$ dependence enters only in the second and higher order
terms; this expression agrees with the perturbation approximation
(\ref{L2}) to first order.

\subsubsection{Proportional noise}
In the case where the noise elements satisfy $b_{ij} = qA_{ij}$
with $q < 1$, we apply \ref{app2} with $p=1$ and proceed as above
to find
\begin{eqnarray*}
L^{UP}_2 &\approx& \ln[ \LL^2(\frac{1 + q^2w^2 + \sqrt{(1 -
q^2w^2)^2 +
4q^2w^4}}{2})] \\
&\approx& L_2^0 + (qw^2)^2 + (qw^2)^4(\frac{1}{w^2} - \frac{3}{2})
+ O(qw^2)^6,
\end{eqnarray*}
where $w^2$ was defined previously (\S  \ref{s:ndependence}) as
$\sum_jv^2_ju_j^2$, and the approximation in the second line is
valid in the limit of small $qw^2$. As expected this agrees with
the perturbation approximation result (\ref{L2}) for proportional
noise to first order.

\subsection{Further approximation}
\label{s:furtheriter} The above treatment is completely accurate
in the way it handles the $\B$ for homogeneous noise. Any
inaccuracy stems from using the approximation $\A^r \approx \LL^r
\uu\vv^T$ on the $\A$ for $r=1$. We can improve on this inaccuracy
to any desired degree, as explained below. Unfortunately, any
approximation past the first order is only applicable to
homogeneous noise (all variances the same) and not to proportional
noise or any other form with different variances. For the
remainder of this section, therefore, only homogeneous noise will
be considered.

\subsubsection{Second approximation} To illustrate the idea,
we begin with a second approximation wherein (\ref{app2}) is
assumed to be accurate for $r=2$ and higher, but not $r=1$. In
this second approximation, $\A$'s which occur ``alone''
(surrounded by two $B$'s) in an element contribute a factor
$\alpha_1 \lambda^2_1$ instead of just $\LL^2$.

We now break $\x^t$ into $\x^t = \x_{AA}^t + \x_{AB}^t + \x_B^t$
analogously to (\ref{xt0}), where $\x_{AA}^t$ are the terms
beginning with $\A\A$, etc. Proceeding just as above, we find that
\[
   \left( \begin{array}{c}
     \lb |\x_{AA}^{t+1}|^2 \rb \\
     \lb |\x_{AB}^{t+1}|^2 \rb \\
     \lb |\x_{B}^{t+1}|^2 \rb
     \end{array} \right) \approx
      \M_2
 \left( \begin{array}{c}
     \lb |\x_{AA}^t|^2 \rb \\
     \lb |\x_{AB}^t|^2 \rb \\
     \lb |\x_B^t|^2 \rb
     \end{array} \right)
\]
with
\[
\M_2 = \left( \begin{array}{ccc}
     \LL^2 & \LL^2/\alpha_1 & 0 \\
     0 & 0 & f_v\alpha_1\LL^2/n \\
     nf_ub^2 & nf_ub^2 & n^kb^2\end{array} \right),
\]
where $f_u$, $f_v$ and $k$ were previously defined in (\ref{f_u}),
(\ref{f_v}) and (\ref{k}) and account for the difference between
UH noise and T noise.

The second moment will diverge when the largest eigenvalue of
$M_2$ is greater than 1. This eigenvalue is the largest root of
the equation
\begin{eqnarray*}
\mu^3 &-& \mu^2(\lambda_1^2 + n^kb^2) +
\mu\lambda_1^2b^2(n^k-f_uf_v\alpha_1) + \\
&& ~~~~b^2\lambda_1^4f_uf_v(1-\alpha_1) = 0.
\end{eqnarray*}
Notice that in the limit $\alpha_1 = 1$, that is, the limit that
the first approximation is accurate, we recover the characteristic
equation for the first approximation (\ref{mu_UH}).

\subsubsection{Higher order approximation}
\label{s:higherorderapproximation} We can extend the above
procedure to any level of accuracy. Define a vector
$\mathbf{\alpha}$ by
\[
\alpha_r = \frac{1}{f_uf_v \lambda_1^{2r}} \sum_{ab} (A^r_{ab})^2
\]
The elements of $\mathbf{\alpha}$ are the successive corrections
to (\ref{app2}). As $r$ increases, $\alpha_r$ tends to 1 because
$\lambda_1^p \gg \lambda_i^p$ becomes very accurate for large $p$.
The Lyapunov exponent of the system is given by the log of the
largest eigenvalue of
\begin{equation}\label{M_r}
\footnotesize
  \M_r = \left( \begin{array}{cccccc}
     \lambda_1^2 & \lambda_1^2\frac{1}{\alpha_r} & 0 & \cdots & 0 & 0 \\
     0 & 0 & \lambda_1^2\frac{\alpha_r}{\alpha_{r-1}} & \cdots & 0 & 0 \\
     \vdots & \vdots & \vdots & \ddots & \vdots & \vdots \\
     0 & 0 & 0 & \ldots & \lambda_1^2\frac{\alpha_2}{\alpha_1} & 0 \\
     0 & 0 & 0 & \ldots  & 0 & \lambda_1^2\alpha_1f_v/n \\
     nf_ub^2 & nf_ub^2 & nf_ub^2 & \ldots & nf_ub^2 & n^kb^2\end{array} \right).
\end{equation}
in the large $p$ limit. The characteristic equation for this
matrix can be expressed iteratively as in \S \ref{s:large_noise}
below, but the largest eigenvalue must be computed numerically.
This method is exact for any noise and any $\A$ with a simple,
dominant eigenvalue, however ill-conditioned $\LL$ may be.

\begin{figure}[hbt]
     \centering
      \includegraphics[width=3in]{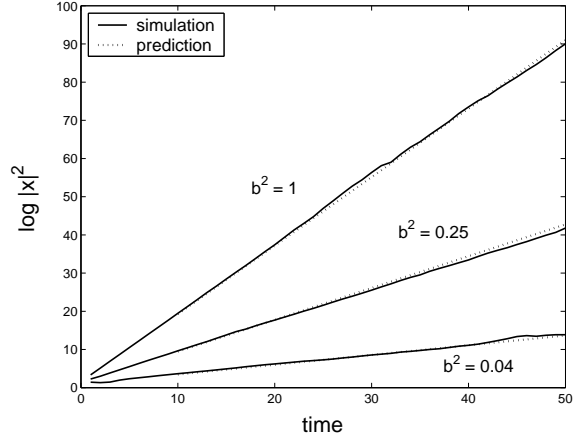}
     \caption{Accuracy of approximation for very ill-conditioned system with large noise}
    \label{f:crazyA}
\end{figure}
The accuracy of the higher-order approximation method is
demonstrated in the simulation of figure \ref{f:crazyA}. In this
figure, the second moment of a system with a very poorly-behaved
$\A$ (equation \ref{crazyA}), subject to normal UH noise with
various $b^2$, is simulated. The solid lines are the average over
10,000 runs of the simulation, and the dotted lines are the
analytical prediction of \S \ref{s:higherorderapproximation} for
$r = 6$. The $\A$ in this system has $v^2 = 170.51$ and so
virtually any noise is not treatable using the perturbation
approximation. The slopes of the dotted lines are 0.255, 0.833,
and 1.794 for $b^2 = 0.04, 0.25$, and 1 respectively. Compare to
the perturbation approximation which estimates 2.02, 3.85 and
5.24. This system is convergent in mean with $\LL = 0.95$; notice
how quickly the moments diverge even for small noise because $v^2$
is large.

\subsection{Large noise limit} \label{s:large_noise} Using the
results of the higher order approximation, we can obtain an
approximation for the Lyapunov exponent for second moment
evolution in the large noise limit.

Note that situations where a small noise has a large effect
because the system's dominant eigenvalue is ill-condi\-tioned
 (\S \ref{s:multivariateproperties}) are \textit{not} treatable using
the formalism of this section, for the reasons given below.

\subsubsection{Criterion for large noise}
A good estimate for the onset of the large noise regime can be
obtained by comparing the largest eigenvalue of $\A$ to the
average largest eigenvalue of the $\B$.

For independent (UH) noises with variance $b^2$ chosen from a
normal distribution, the magnitude of the largest eigenvalue of
$\B$ is given on average by \cite{Geman_largest_eigenvalue}
\begin{equation}\label{largest_eig_normal_UH}
  \overline{\LL_B} = b\sqrt{n}, \; \mbox{ independent noises}
\end{equation}
For correlated (T) noises chosen from a normal distribution, the
matrix $\B$ is simply a normal random multiple of the matrix $\G$
of all ones. $\G$ has largest eigenvalue $n$ and so the largest
eigenvalue of $\B$ is on average
\begin{equation}\label{largest_eig_normal_T}
  \overline{\LL_B} = bn, \; \mbox{ correlated noises}
\end{equation}

The large noise case corresponds to $\LL_B \gg \LL$, that is,
\begin{equation}\label{large_noise_condition}
  n^k b^2 \gg \LL^2
\end{equation}
where $k=1$ for independent (UH) noise and $k=2$ for correlated
(T) noises.

\subsubsection{Lyapunov exponent}
To find the Lyapunov exponent for second moment evolution in the
large noise limit that $n^k b^2 \gg \LL^2$, we introduce the small
parameter
\begin{equation}\label{large_noise_delta}
  \delta = \frac{\LL^2}{n^kb^2} \ll 1.
\end{equation}
We will appeal to the iteration treatment of  \S
\ref{s:higherorderapproximation} which was accurate for any noise.
Recall that the Lyapunov exponent is given in this treatment by
the log of the largest eigenvalue $\mu_r$ of the matrix $M_r$
(equation \ref{M_r}), where $r$ is any integer. The approximation
is more accurate the larger $r$ is, but as we will see, in the
large noise limit there is no need to consider large $r$.

Note that the parameters $\{\alpha_i\}$, $f_u$, and $f_v$ enter
into the calculation of $\mu_r$ (equation \ref{M_r}). If these
parameters are large, they can ruin our expansion since they
multiply $\delta$. We therefore assume that they are $O(1)$. Note
that this amounts to assuming the system is well-behaved; this is
why, as noted above, this expansion is not applicable to
ill-conditioned systems.

The characteristic equation for $\M_{r+1}$ can be written
\[
0 = (\LL^2 - \mu) [\mu D_r + \alpha_r\LL^{2r}b^2f_uf_v] +
\LL^{2r}b^2f_uf_v
\]
where $\mu$ are the eigenvalues and the $D_r$ are defined
recursively by
\[
D_{r+1} = -\mu D_{r} + \alpha_r \LL^{2r}b^2f_uf_v
\]
with
\[
D_2 = \det \left( \begin{array}{cc}
     -\mu & f_v\alpha_1\LL^2/n \\
     nf_ub^2 & n^kb^2 - \mu \end{array} \right).
\]

Keeping only terms to first order in $\delta$ in the
characteristic equation above, we find that the largest eigenvalue
$\mu_{r+1}$ is given approximately by
\begin{equation}\label{large_noise_mu_r}
  \mu_{r+1}^2 = n^kb^2\left( \frac{1 + \delta + \sqrt{(\delta - 1)^2 +
\delta \frac{4\alpha_1f_uf_v}{n}}}{2} \right)
\end{equation}
independent of $r$, so that
\begin{eqnarray}\label{iter_approx_L2_large_noise}
L_2 &\approx& \log n^kb^2 + \delta\frac{\alpha_1f_uf_v}{n^k} +
O(\delta^2) \\
&\approx& \log n^kb^2 + \LL^2\frac{\alpha_1f_uf_v}{n^{2k}b^2} +
O(\LL^4).
\end{eqnarray}
Note that the zeroth order term corresponds to that found in the
$\A=0$ limit by different means in \S \ref{s:Aequal0}.

\begin{figure}[hbt]
     \centering
      \includegraphics[width=3in]{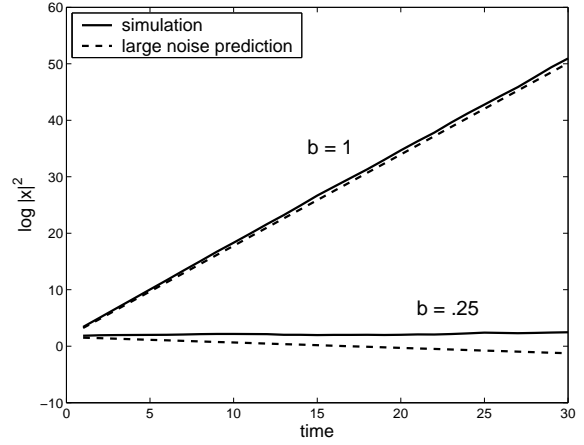}
     \caption{Accuracy (and inaccuracy) of large noise prediction}
    \label{f:large_noise}
\end{figure}
The range of applicability of the large noise approximation is
demonstrated in figure where the second moment is plotted against
the prediction of (\ref{iter_approx_L2_large_noise}). The $\A$ of
(\ref{exA}) is used, which has $n=5$ and $\LL = 0.96$. Thus $b=1$
is well within the large noise regime, as shown in the figure.
Even $b = 0.25$, which is not within the large noise regime, is
reasonably well predicted by this approximation.

\section{Critical value and stability diagram} \label{s:bcrit}
As a general rule, large deviations from the average become
reasonably likely when the noise is large enough that the second
moment diverges. The onset of second moment divergence therefore
marks a threshold between two types of behavior and defines a
critical value of the size of the noise.

In the case of a homogeneous noise, in which all the noise
elements have the same variance $b^2$, the critical value can be
simply expressed as the value of the variance where the second
moment Lyapunov exponent $L_2$ equals 1. For a proportional noise,
the critical value is the value $q_c$ of the constant of
proportionality (see table \ref{t:correlationrules}) for which
$L_2=1$. The approaches used in this paper allow for a detailed
treatment of the critical value $b_c^2$ in the case of homogeneous
noise, but unfortunately not for proportional or other forms of
noise, which is left as a topic for future work. Proportional
noise is only discussed in the mean value case where it takes the
same form as homogeneous noise.

Throughout this section we will assume that the system is
well-behaved, that is, $f_u$, $f_v$ and the $\{\alpha_i\}$ are
close to 1. Recall that $f_u$ and $f_v$ account for the difference
between independent and correlated noises, and the $\{\alpha_i\}$
measure the accuracy of the approximation (\ref{app2}) for
successive powers $\A^i$.

\subsection{$\LL^2 \rightarrow 1$ limit}
\label{s:crit_small_noise} A limit of particular interest when
considering the critical value is $\LL^2 \rightarrow 1$, where, as
we will see, the critical value drops sharply to 0. Since only
small noise is required to cause divergence in this limit,  we can
apply the first approximation of the iteration treatment, in
particular equation \ref{mu_UH}, to find the critical value
\begin{equation}\label{bcritu_small}
  (b_c^2)_{\mbox{\footnotesize small noise}} \approx \frac{1}{n^k + \frac{f_uf_v\LL^2}{1-\LL^2}}.
\end{equation}
where $k = 1$ for independent noises and $k=2$ for correlated
noises. The sharp dropoff to 0 of the critical value is evident
from this expression and demonstrated in figure \ref{f:phaseplotb}
below. Note that when $n=1$, in which case $f_u=f_v=1$ and $\LL =
a$ we retrieve the scalar result $b^2_c = 1 - a^2$.

Our above expression for $b_c^2$ should coincide with that implied
by the perturbation approximation of \S \ref{s:perturbation}.
Using equation \ref{L2} combined with table
\ref{t:varepsilonnoises} and definitions (\ref{f_u}) and
(\ref{f_v}) for $f_u$ and $f_v$, we obtain the perturbation
approximation result
\[
(b_c^2)_{\mbox{\footnotesize pert}} = \frac{1-\LL^2}{f_uf_v},
\]
which is easily seen to be equivalent to (\ref{bcritu_small}) to
first order in $(1-\LL^2)$.

\subsection{$\LL \To 0$ limit}\label{s:crit_large_noise}
When $\LL^2$ is small and the system well-behaved, the second
moment will only diverge if the noise is large. In this case the
limit can be obtained from the large noise treatment of \S
\ref{s:large_noise}, in particular equation
(\ref{large_noise_mu_r}) whence we find
\begin{equation}\label{bcritu_large}
  (b_c^2)_{\mbox{\footnotesize large noise}} \approx \frac{1}{n^k +
  \frac{\alpha_1f_vf_u\LL^2}{1-\LL^2}}.
\end{equation}
For well behaved systems with $\alpha_1 \approx 1$, this
expression is almost equivalent to the small noise expression
(\ref{bcritu_small})! The scalar result is again retrieved from
this expression. We also note that when $\LL = 0$, we obtain the
critical value
\[
(b_c^2)_{\LL = 0} = 1/n^k,
\]
which was obtained in a different way in section \ref{s:Aequal0}.

\subsection{Stability diagram} \label{s:stability_diagram}
In the case of a well-behaved system with $\alpha_1 \approx 1$,
the functional form of the critical value is the same for small
and large values of $\LL^2$. We thus propose the following
expression
\begin{equation}\label{bcrit}
  b_c^2 = \frac{1}{n^k +
  \frac{f_vf_u\LL^2}{1-\LL^2}}
\end{equation}
for the critical value for all ranges of homogeneous noise in
well-behaved systems.
\begin{figure}[htb]
      \centering
      \includegraphics[width=3in]{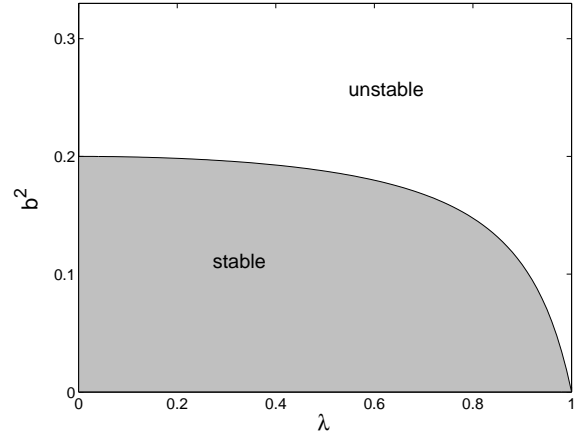}
     \caption{Second moment stability diagram for $n=5$}
    \label{f:phaseplotb}
\end{figure}
Using this expression we can make a phase plot for the stability
regions of the system, as shown in figure \ref{f:phaseplotb} for
$n=5$ and independent (UH) noise.

As for proportional noise, we can produce a stability diagram to
compare with the homogeneous noise case in the mean value limit.
This diagram indicates how large, as a fraction of the size of the
unperturbed elements, the noise must be to cause divergence.

Recall that the size of a proportional noise was defined by the
factor $q$, the constant of proportionality between the typical
noise size and the average element of $\A$ (see table
\ref{t:correlationrules}). In the mean value approximation, the
largest eigenvalue is given approximately by $\LL \approx an$ and
we thus obtain the critical value
\[
q_c = \frac{n}{\LL} b_c .
\]
This relation produces the phase plot of figure \ref{f:phaseplotq}
for the constant of proportionality $q_c$ as a function of $\LL$.
\begin{figure}[htb]
      \centering
      \includegraphics[width=3in]{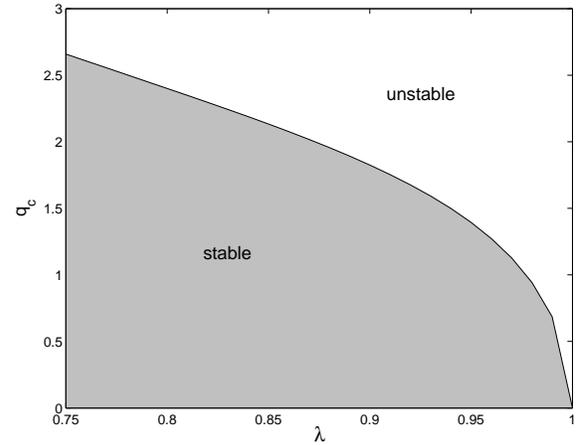}
     \caption{Stability diagram for proportional noise in mean value approximation, $n = 5$}
    \label{f:phaseplotq}
\end{figure}

\subsection{$n$ dependence of critical
value}\label{s:crit_n_dependence} Expression (\ref{bcrit}) can be
used to study the $n$ dependence of the critical value. The $n$
dependence is weak for large $\LL$, but it is strong when $\LL$ is
small and a large noise is required to cause divergence.

In the limit $\LL \To 1$, where only a small noise is needed to
create second moment divergence, it is seen from expression
(\ref{bcrit}) that the $n$ dependence is quite weak. Indeed, the
expansion (\ref{iterapproxL2}) for small noise showed that $n$
dependence enters only in the second order term in this limit. The
effects of different forms of noise in this case are quite
similar.

\begin{figure}[hbt]
     \centering
     \includegraphics[width=3in]{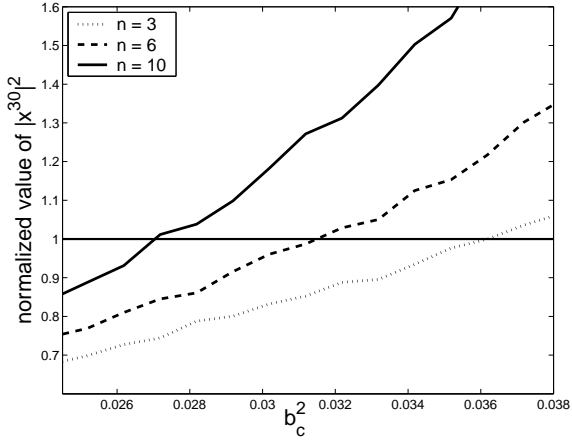}
     \caption{$n$ dependence of critical value}
    \label{f:bcritndependence}
\end{figure}
The $n$ dependence of the critical value for small, independent
(UH) noises is demonstrated in figure \ref{f:bcritndependence}.
The figure plots the value of the second moment at $t=30$ for
systems with various $b$, normalized and averaged over 300,000
runs. Because of the normalization, the initial value (at $t=0$)
of the system's second moment was 1. The critical value $b_c^2$ is
thus indicated for each $n$ by $x$-coordinate for which the
plotted curve's $y$-coordinate begins to exceed 1. Compare the
values of $b^2_c$ given by the simulation to the analytic
estimates $b^2_c = 0.0365, 0.0326, 0.0288$ for $n = 3, 6, 10$,
respectively from equation (\ref{bcrit}). The values of the
elements of the $\A$s used in this plot were generated randomly
from a normal distribution\footnote{the matrix was not accepted if
it did not have a simple dominant eigenvalue}, and the matrices
were normalized to have $\LL = 0.98$.

In the opposite limit of $\LL \To 0$ and large noise, the $n$
dependence is quite strong, as shown in
(\ref{iter_approx_L2_large_noise}). It is in this limit that the
difference between independent and correlated noises is quite
marked, due to the ``destructive interference'' phenomenon of
independently varying noises discussed previously in \S
\ref{s:ndependence}.

\subsection{Comparison to convergence bounds}
\label{s:crit_compare_to_bounds} The critical value (\ref{bcrit})
provides a much more accurate estimate of the ``safe'' level of
noise for which the second moment does not diverge than do the
convergence bounds of appendix \ref{s:bounds}. We bring this point
up  because the traditional mathematical approach to stochastic
stability is to use bounds.

These bounds are $b^2 < \frac{(1-\lambda_1^2)}{n}$ for the second
moment (\ref{convcondlim}) and $b^2 < \frac{(1-\lambda_1)^2}{4n}$
for any moment (\ref{normconvlim}) in the large $n$ limit. For
typical well--conditioned systems with $v^2$ relatively close to 1
(figure \ref{f:vgap}, appendix \ref{s:conditionetc}), it is clear
that the critical value is much less restrictive than either of
the bounds. That is, the bounds stipulate that we must take a very
small noise to guarantee convergence of the second moment; but the
critical value indicates that the second moment will converge for
a much larger range of noise. When $v^2$ is large, $\LL$ and thus
the matrix $\A$ are ill-conditioned and the norm of $\A$ is
typically much greater than $\LL$, so the above bounds are not
accurate.

\section{Results are generally inapplicable to continuous limit}
\label{s:continuouslimit} As a last subject, we discuss the
continuous limit of our stochastic system. The results of this
paper are not generally applicable in the continuous limit because
there is no such thing as small noise, in the sense we have used,
in the continuous limit. Of the cases we have considered, the
discrete result is only applicable to the continuous limit in the
mean value approximation that $\A \approx a\G$.

The reason that the relative size of the noise depends on the time
scale is that the correct limit of a white noise process has
standard deviation proportional to $\sqrt{dt}$
\cite{KloedenPlaten}. Thus, the noise necessarily dominates as $dt
\rightarrow 0$. To illustrate this point, consider a particle
moving in a one dimensional diffusion process
\[
dx = ((a-1) dt + b dw)x
\]
where $dw$ is a Wiener process. When we consider the system's
average motion on a large time scale, the particle generally
progresses along the curve $x_0e^{(a-1)t}$. However, on very small
time scale, the motion is completely erratic because it is
dominated by the noise.

For multidimensional systems, the continuous limit of
(\ref{main1}) is the stochastic differential equation (in the Ito
sense)
\begin{equation}\label{sde}
  d\x = [(\A - \I)dt + d\B]\x
\end{equation}
where $d\B = bd\W$ is a matrix of Wiener processes with mean 0 and
standard deviation proportional to $b\sqrt{dt}$, and $\I$ is the
identity matrix. Again, in the $dt \rightarrow 0$ limit, the
motion is completely dominated by the noise and the vector $\x$ is
transformed erratically around in $\mathbf{R}^n$. The system can
never become aligned with $\uu$ because the large noise causes it
to couple with the other modes of $\A$. Only when $\lambda_2
\rightarrow 0$ does the system become aligned with $\uu$ and
behave similarly to the perturbation approximation, above.

\subsection{Correspondence between continuous and discrete results
in the mean value approximation} \label{s:mvacontinuous} For
correspondence between the discrete and continuous cases we
consider a system in which $\lambda_2 = 0$: the mean value limit
that $\A = a\G$. For this $\A$ and totally correlated noise, an
analytic solution to (\ref{sde}) is possible because $\A$ and $d\W
= \G dw$ commute \cite{Oksendal,Mao}. Here $dw$ is a
one-dimensional Wiener process and $\LL = na$ is the only nonzero
eigenvalue of $\A$. The solution to (\ref{sde}) is
\[
\x(t) = e^{(\A- \I - b^2\G^2/2)t + b\G w}\x(0).
\]
where $w = \int_0^t dw$ is normal with variance $t$. From this it
is straightforward to calculate that
\[
 \lb |\x(t)|^p \rb=
e^{pt[(na - 1) - \frac{n^2b^2}{2}]}e^{\frac{b^2p^2n^2}{2}t}
\left|\frac{1}{n}\G\x(0) \right| ^p
\]
in the asymptotic limit, and the moment Lyapunov exponent is
\[
\ell_p = -p\delta + p(p-1)\frac{n^2b^2}{2}
\]
where we have taken $\LL = na = 1 - \delta$. This can be compared
with the discrete result for the mean value limit and totally
correlated noise:
\[
L_p \approx -p\delta + p(p-1)\frac{n^2b^2}{2}(1 - 2\delta) +
O(\delta^2) + O(b^4),
\]
where we have applied $v_i = u_i = n^{-1/2}$. In the limit of
small time step the expressions are equivalent to lowest order.
This same analysis can also be performed for a scalar system where
there are no other modes to couple to.

\subsection{Failure of discrete result in the continuous
limit} \label{s:arnold} When there are nonzero modes for $\A$
other that $\LL$, the discrete result should not, and does not,
correspond to the continuous limit. This can be verified by
comparison to the result of \cite{Arnoldperturbation} for small
noise moment Lyapunov exponents of arbitrary two-dimensional
linear stochastic differential equations. This result, for white
noise, is
\begin{equation}\label{arnold}
  \ell_p = -p\delta + p\gamma_1\frac{b^2}{2} + p^2\gamma_2\frac{b^2}{2} +
pO(b^2) + O(p^2)
\end{equation}
where we take $\LL = 1 - \delta$. The $\gamma$ factors depend on
the form of noise considered. $\gamma_2$ depends only on the
dominant eigenmode, while $\gamma_1$ depends on both eigenmodes.
To proceed we assume UH noise for definiteness, wherein one can
show that $\gamma_2 = v^2$. The discrete result (\ref{L2}) for UH
noise is
\begin{equation}\label{cont2dimUHnoise}
  L_p \approx -p\delta + p(p-1)\frac{v^2b^2}{2}(1 - 2\delta) + O
(\delta^2) + O(b^4).
\end{equation}
Comparing this expression to the continuous version
(\ref{arnold}), we see that the $1-2\delta$ factor on the noise
term accounts for the difference between discrete and continuous
evolution, as in \S \ref{s:mvacontinuous}, and that the
$p^2\gamma_2\frac{b^2}{2} + pO(b^2)$ in (\ref{arnold}) probably
corresponds to the $p(p-1)\frac{v^2b^2}{2}$ term in
(\ref{cont2dimUHnoise}). Note that the $p(p-1)$ form is present in
both continuous scalar and T noise cases and is typical of
log-normal distributions.

However, the term in (\ref{arnold}) proportional to $\gamma_1$ is
completely absent in the discrete result; moreover, it depends on
$\lambda_2$ and its eigenvector which have no effect on the small
noise discrete system. This term shows how the solution is coupled
to all modes, not just the dominant one, in the continuous limit.
In fact, for UH noise, one can show (see (\ref{v2})) that
$\gamma_1 = 1 - v^2$; in the mean value limit $v^2 = 1$ and the
contribution of the second mode is 0.


\section*{Acknowledgement}
The author would like to thank David Luenberger, Gene Golub, and
especially Rob Schreiber and Bernardo Huberman for helpful
discussions and suggestions. This work was partially supported by
the National Science Foundation under Grant No. 9986651.


\appendix

\section{Matrices used to generate figures}
All of the unperturbed matrices $\A$ used to generate the figures
of this paper were randomly generated. Two particular $\A$s are
presented here. The others were generated as described in the text
from distributions with low variance, and it would be a waste of
space to present them exactly (see discussion in \S
\ref{s:Aproperties}).

The matrix which was used to generate figure \ref{f:fluct} and
many others as noted in the text is
\begin{equation}\label{exA}
\footnotesize \A = \left( \begin{array}{ccccc}
    0.1795  &  0.0861  &  0.1860  &  0.0924  &  0.1661 \\
    0.1429  &  0.1680  &  0.0517  &  0.2626  &  0.3272 \\
    0.3558  &  0.0127  &  0.2797  &  0.0221  &  0.3227 \\
    0.2766  &  0.2654  &  0.1611  &  0.0408  &  0.0745 \\
    0.3539  &  0.3059  &  0.0596  &  0.2933  &  0.3147 \\
    \end{array} \right).
\end{equation}
This matrix has largest eigenvalue $\LL = 0.966$, second largest
eigenvalue $|\lambda_2| = 0.228$, and $v^2 = 1.10$ as computed by
Matlab. This matrix is thus quite well-behaved as defined in \S
\ref{s:Aproperties}.

The matrix with ill-conditioned $\LL$ used to generate the plot of
figure \ref{f:crazyA} is
\begin{equation}\label{crazyA}
\footnotesize \A = \left( \begin{array}{ccccc}
    0.5086  &  0.3496  &  0.0795 &  -0.2044  & -0.3530 \\
   -0.6168  &  0.1553  &  0.5224 &  -0.0293  &  0.0137 \\
   -0.5526  &  0.0069  &  0.0008 &  -0.3189  &  0.4345 \\
    0.4805  &  0.8053  & -0.5502 &   0.6173  & -0.3041 \\
   -0.4307  &  0.8960  &  0.0255 &   0.1454  &  0.6965 \\
    \end{array} \right)
\end{equation}
with largest eigenvalue $\LL = 0.950$, second largest eigenvalue
$|\lambda_2| = 0.888$, and $v^2 = 170.3$ as computed by Matlab.

\section{Reduction of nonnegative stability analysis to primitive
systems} \label{s:primitive} The reason that $\LL$ is simple and
dominant in all nonnegative systems of interest is that we need
only consider systems with primitive $\A$, and primitive matrices
have the above property by the Perron-Frobenius theorem. Stability
analysis of any nonnegative system whose matrix is not primitive
reduces to analysis of primitive subsystems.

More precisely, nonnegative matrices which are not primitive may
be either reducible or irreducible imprimitive. Reducible matrices
are those which can be written in the form
\begin{equation}\label{redform}
  \left(
\begin{array}{cc}
            \mathbf{C} & \mathbf{X} \\
       \mathbf{0} & \mathbf{D} \end{array} \right),
\end{equation}
where $\mathbf{C}$ and $\mathbf{D}$ are square, by renaming the
indices\cite{BermanPlemmons}. Stability analysis reduces to
analysis of the subsystems $\mathbf{C}$ and $\mathbf{D}$, because
$\left(
\begin{array}{cc}
            \mathbf{C} & \mathbf{X} \\
        \mathbf{0} & \mathbf{D} \end{array} \right)^n = \left( \begin{array}{cc}
            \mathbf{C}^n & \mathbf{Y} \\
        \mathbf{0} & \mathbf{D}^n \end{array} \right)$. A similar
reduction occurs on the subsystems unless they are irreducible.
Irreducible imprimitive  matrices can be written as
\begin{equation}\label{imprimform}
  \left(
\begin{array}{ccccc}
        \0 & \mathbf{C}_{12} & \0 & \ldots & \0 \\
        \0 & \0 & \mathbf{C}_{23} & \ldots & \0\\
        \vdots & & \vdots & \ddots & \vdots\\
        \0 & \0 & \0 & \ldots & \mathbf{C}_{h-1 h} \\
        \mathbf{C}_{h1} & \0 & \0 & \ldots & \0
         \end{array} \right),
\end{equation}
where the 0 blocks along the diagonal are square (second part of
Perron Frobenius theorem). The $h$th power of such a matrix is
block diagonal and the blocks are primitive \cite{BermanPlemmons},
so the stability analysis is again reduced.

Physically, primitive matrices have the property that their powers
are positive\footnote{More exactly, the $p$th power of a
nonnegative $n \times n$ primitive matrix $\A$ has no zero
elements for all $p \geq \gamma(\A)$, where $\gamma(\A)$ (the
index of primitivity) is at most $n^2 - 2n + 2$, and usually much
less\cite{BermanPlemmons}).} (have no 0 elements). From a physical
perspective, primitive systems are thus ``fully interacting''.
This is in contrast to other nonnegative matrices which have zero
blocks when raised to any power.

\section{Further discussion of properties of $\A$}
\label{s:conditionetc} There is a correlation between an
ill-conditioned $\LL$ and a small eigenvalue gap. This is so
because a matrix with a large $\kappa(\LL)$ is close to a matrix
where $\LL$ is repeated. In particular\cite{Golubbook}, there
exists a matrix $\mathbf{E}$ such that $\LL$ is a repeated
eigenvalue of $\A + \mathbf{E}$ and
\[
|\mathbf{E}| \leq \frac{|\A|}{\sqrt{(\kappa(\LL))^2 - 1}}.
\]
However, $\kappa(\LL)$ may be small even if the gap is small. The
relation between $\kappa(\LL)$ and the eigenvalue gap is shown in
figure \ref{f:vgap}, above.

There is also a correlation between normality of $\A$ and a small
$\kappa(\LL)$. When $\A$ is normal, that is, $\A\A^T = \A^T\A$,
all of its eigenvectors are orthogonal and all the eigenvalues are
perfectly conditioned. However, $\kappa(\LL)$ may be small in
matrices which are far from normal. The relation between
$\kappa(\LL)$ and the normality of $\A$ is shown in figure
\ref{f:hen}.

\begin{figure*}[hbt]
     \centering
      \includegraphics[width=3in]{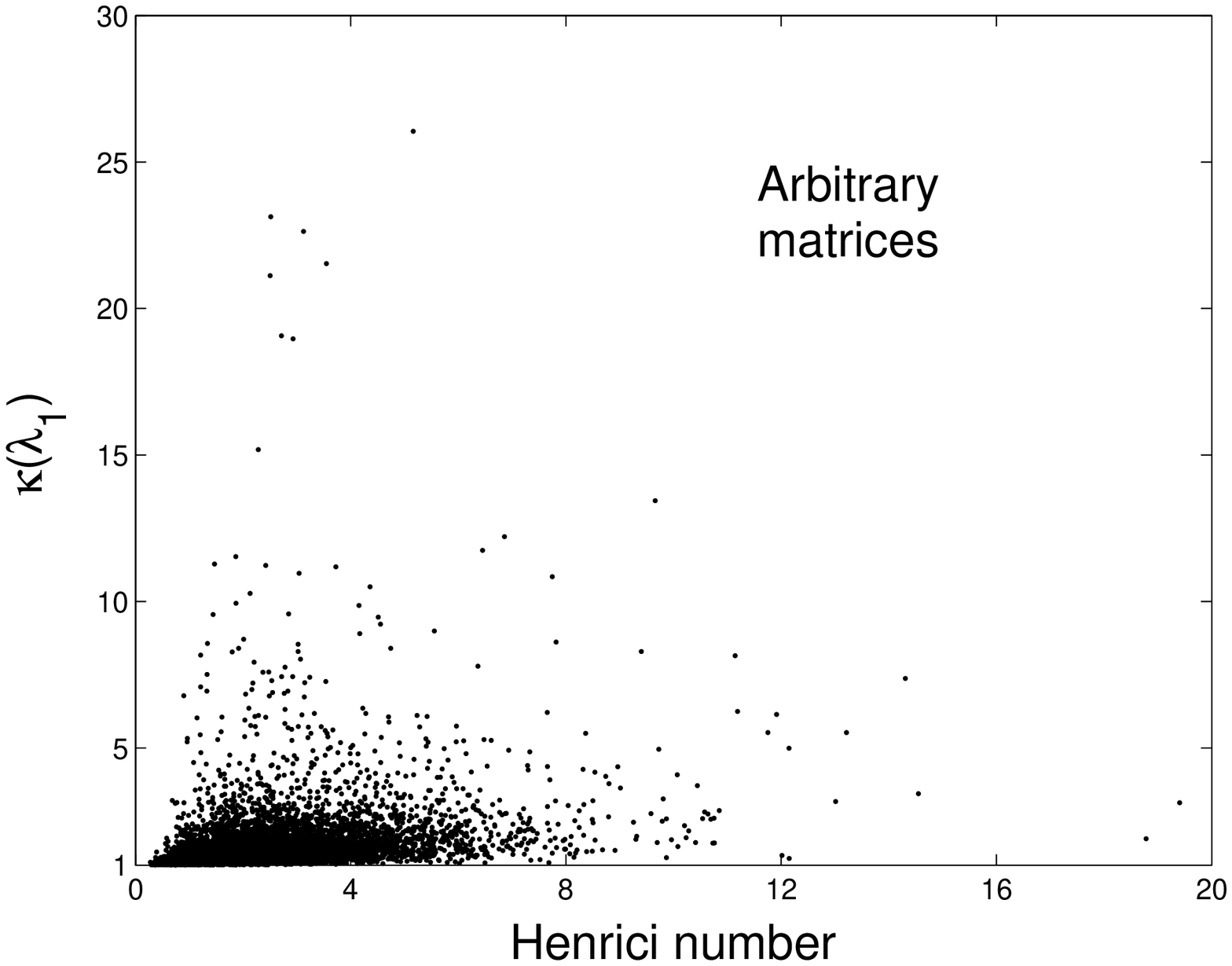}
      \includegraphics[width=3in]{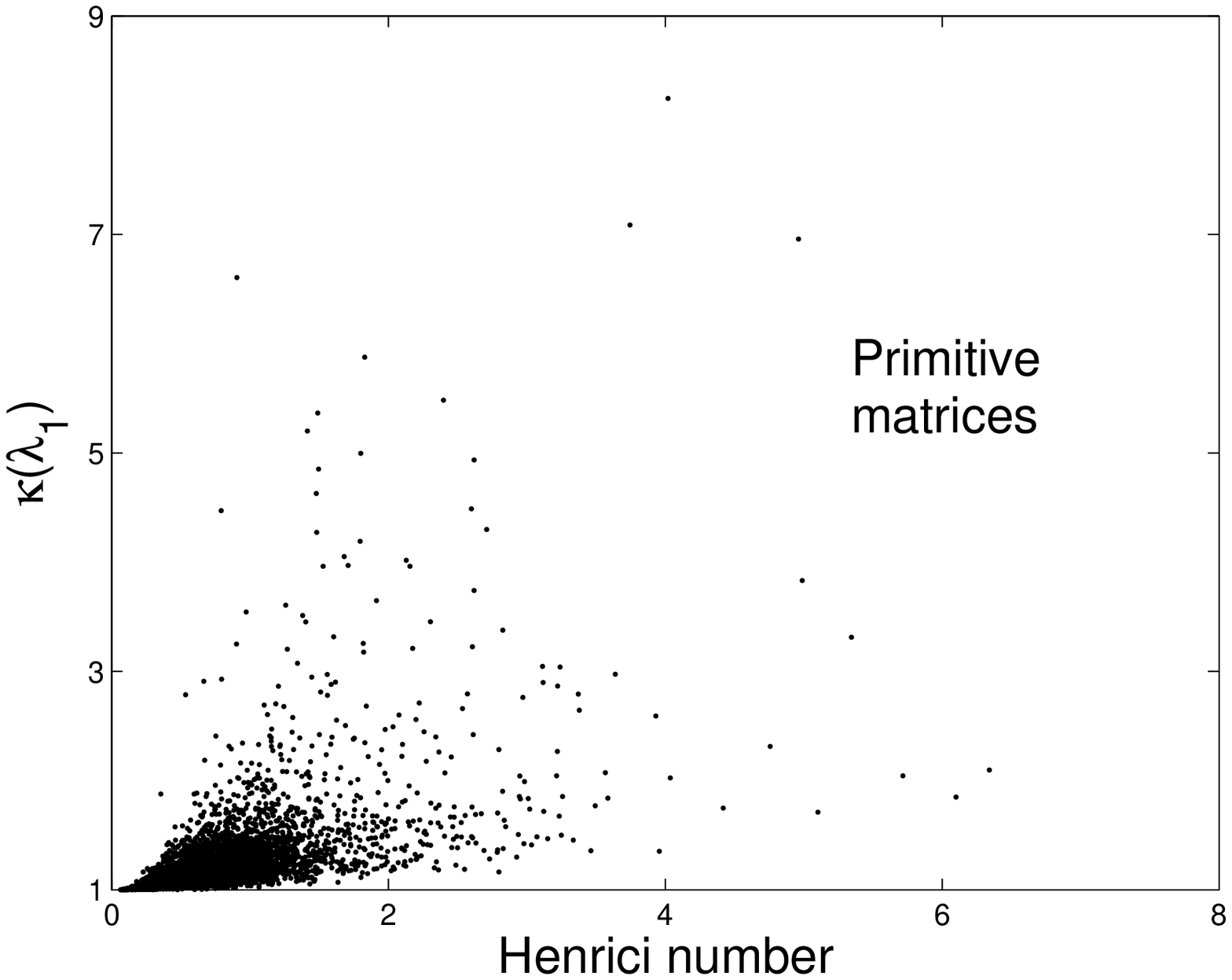}
     \caption{Scatter plots of $|\vv|$ versus
     the Henrici number $|\A\A^T - \A^T\A|$, a measure of non-normality, for 10,000 randomly generated $5 \times 5$
     matrices normalized so that $\LL = 1$.  At left the elements were chosen from
     a normal distribution and only plotted if $\LL$ was real.
     At right, the matrices are nonnegative primitive; a random number of elements were 0, and the
     nonzero elements were chosen from a uniform distribution.}
    \label{f:hen}
\end{figure*}

Another way to characterize $\kappa$ is the relation
\begin{equation}\label{v2}
  v^2 = 1 - \sum_{i \neq 1} (\uu \cdot \e^R_i) (\vv \cdot \e^L_i)
\end{equation}
where $\e^R_i$ is the $i$th column of $\PPP$ (the right
eigenvector corresponding to $\lambda_i$) and $\e^L_i$ is the
$i$th row of $\Q$ (the left eigenvector corresponding to
$\lambda_i$) This relation is established by noting that
$\sum_{ij}\e^R_i\e^R_j \e^L_i \e^L_j = 1$. It shows how $v^2$ is
related to the angles between the eigenvectors. In particular, we
see that for a normal matrix where the eigenvectors are
orthogonal, $v^2 = 1$; but in general, the angular distribution of
the eigenvalues is complicated.

Finally, there is a correlation between $|\lambda_2| \rightarrow
0$ and the variance $\sigma_A^2$ of the elements of $\A$. Bounds
for the second largest eigenvalue can be found in the case of row
(or column) stochastic matrices, for example\cite{BermanPlemmons}:
$|\lambda_2| \leq \min\big(1 - \sum_i\min_j A_{ij},\sum_i\max_j
A_{ij} - 1 \big)$. This shows that, at least for stochastic
matrices, a small variance $\sigma_A^2$ corresponds to a large
eigenvalue gap. This is shown to be true for all matrices in
figure \ref{f:gap}. Of course, the converse is not true; matrices
with large $\sigma_A^2$ can also have a large eigenvalue gap, as
also is shown in figure \ref{f:gap}.
\begin{figure*}[hbt]
     \centering
      \includegraphics[width=3in]{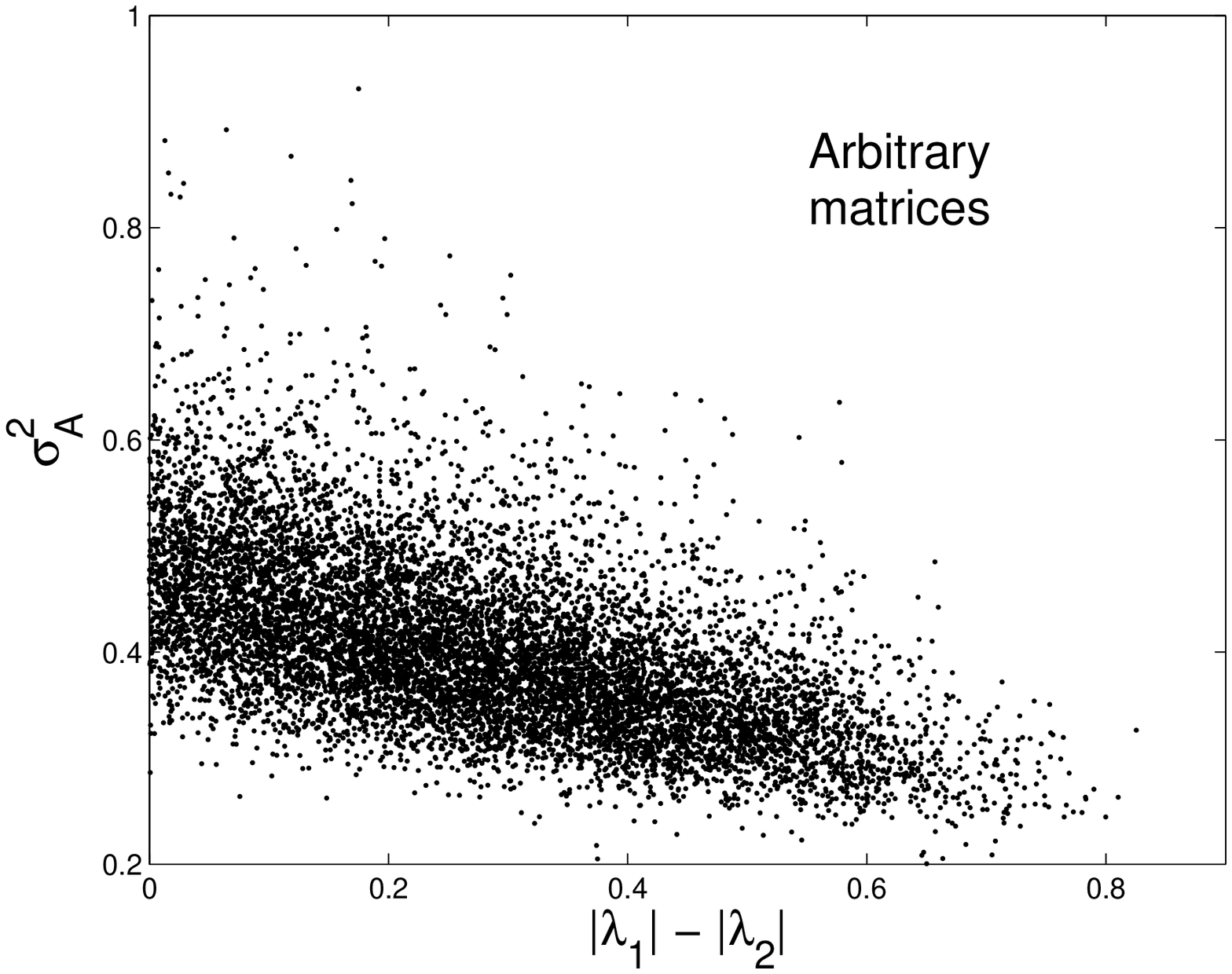}
      \includegraphics[width=3in]{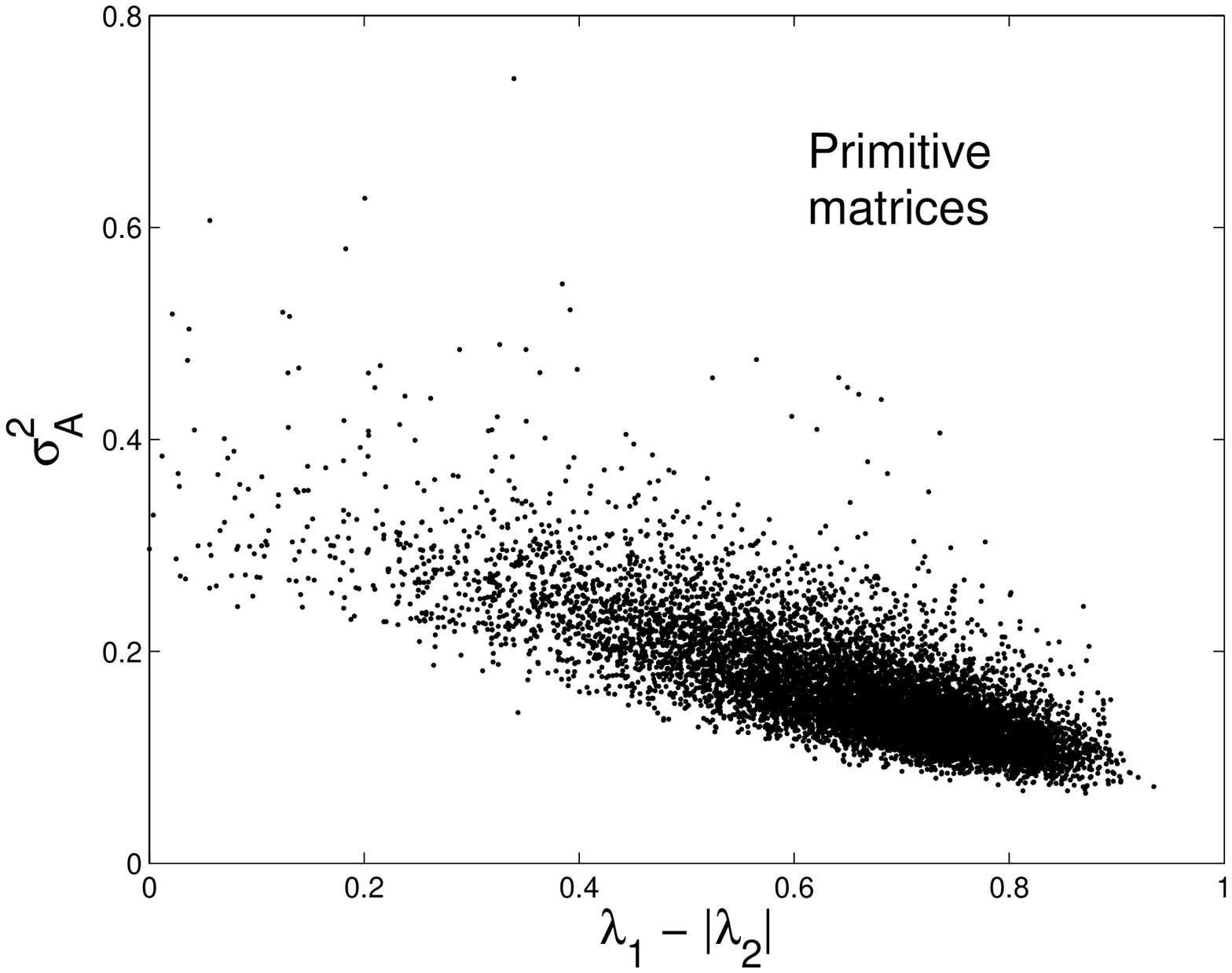}
     \caption{Scatter plots of eigenvalue gap $\LL - |\lambda_2|$ versus
     standard deviation of the $\A_{ij}$ for 10,000 randomly generated $5 \times 5$
     matrices, normalized so that $\LL = 1$. At left the elements were chosen from
     a normal distribution and only plotted if $\LL$ is real.
     At right, the matrices are nonnegative primitive; a random number of elements were 0, and the
     nonzero elements were chosen from a uniform distribution.}
    \label{f:gap}
\end{figure*}

\section{Bounds on convergence of $\lb |\x|^2 \rb$} \label{s:bounds}
In this section we apply the matrix 2-norm to determine two
different bounds on the variance of the noise which, if satisfied,
ensure the convergence of $\lb |\x|^2 \rb$. These conditions are
sufficient but by no means necessary. The second moment will of
course never converge if the system does not converge in mean. We
therefore take $\LL < 1$ in this section.

The norm of a matrix is any function satisfying the regular
properties of a vector norm and additionally the inequality
$|\A\B| \leq |\A||\B|$. The matrix 2-norm corresponding to the
usual Euclidean vector norm is
\begin{equation}\label{2norm}
  |\A|_2 = (\rho(\A\A^{*}))^{1/2}.
\end{equation}
where $\rho$ is the spectral radius. Note that for any norm, $|A|
\geq |\x\A|/|\x| = |\lambda|$ for any eigenvalue $\lambda$, so
that in particular,
\begin{equation}\label{srnorm}
  \LL \leq |\A|.
\end{equation}
For ill-conditioned matrices, which includes those with
ill-conditioned $\LL$, $|\A|$ is typically much larger than $\LL$
\cite{Golubbook}.

\subsection{Bound on convergence of any moment}
We have $|\x^{t+1}|^p \leq {|\A + \B^t|}^p|\x^t|^p$, so that
\[
\lb |\x^t|^p \rb \leq \left[ \prod_{\tau=1}^{t-1} \lb |\A +
\B^\tau| \rb ^p \right] |\x^0|^p
\]
where the expected value goes inside the product because the noise
is white noise. $\lb |\x|^p \rb$ will thus converge for any $p$
provided that $\lb |\A + \B| \rb < 1$ (we neglect the time
superscript because the noise is stationary), or more usefully
\begin{equation}\label{normconv}
   \lb |\B| \rb < 1 - |\A|.
\end{equation}
using $|\A + \B| \leq |\A| + |\B|$. Since convergence of every
moment is a much stronger condition than convergence of just the
second moment, this bound is typically poor when applied to the
second moment.

We may estimate a lower limit for this bound for well-conditioned
systems in the large $n$ limit when the noise is UH (uncorrelated
$B_{ij}$ all with the same variance $b^2$). We do so by using
$|\A| > \LL$ (equation \ref{srnorm}) and a result of \cite{Geman}
that $\liminf |\B| \geq 2b\sqrt{n}$ almost surely in the large $n$
limit, provided that the elements of $\B$ are mean 0 i.i.d. and
their moments do not grow too fast (which is satisfied for any
reasonable noise). Thus
\[
\lambda_1 + 2b\sqrt{n} \leq |\A| + \lb |\B| \rb
\]
and the condition (\ref{normconv}) on $b$ for convergence at least
weaker than
\begin{equation}\label{normconvlim}
  b^2 < \frac{(1-\lambda_1)^2}{4n}
\end{equation}
in the large $n$ limit. That is to say, (\ref{normconv}) is more
restrictive on $b$ than (\ref{normconvlim}). For ill-conditioned
systems, (\ref{normconvlim}) may not be accurate because $|\A|$
may be much larger than $\LL$.

\subsection{Second moment bound}
A different bound on the convergence of the second moment in the
case of UH noise can be found by applying the expected value
before taking norms. We have
\begin{eqnarray*}
\lb |\x^t|^2 \rb &=& \lb |(\A + \B^t)\x^{t-1}|^2 \rb \\
&\leq& (|\A|^2 + nb^2)\lb |\x^{t-1}|^2 \rb
\end{eqnarray*}
where we have used the properties of the norm and the fact that
the noise is UH, white and has mean 0. We thus have the
convergence condition $|\A|^2 + nb^2 < 1$, or
\[
  b^2 < \frac{(1-|\A|^2)}{n}
\]
for the convergence of $\lb |x|^2 \rb$. Note that this condition
is at least weaker than the condition
\begin{equation}\label{convcondlim}
 b^2 < \frac{(1-\lambda_1^2)}{n}
\end{equation}
because of (\ref{srnorm}). Again, (\ref{convcondlim}) may not be
accurate for ill-conditioned $\A$ because $|\A|$ may be much
larger than $\LL$.



\bibliographystyle{h-physrev3}
\bibliography{main}

\end{document}